\documentclass{article}

\usepackage{rasmus2,natbib,graphicx,color}

\def\bx{{\bf x}}

\def\by{{\bf y}}
\def\bz{{\bf z}}
\def\bX{{\bf X}}
\def\bY{{\bf Y}}
\def\bZ{{\bf Z}}

\def\per{\mathrm{per}}

\newcommand{\xn}{{x_1,\ldots,x_n}}

\newcommand{\jmret}[1]{{#1}}

\begin{document}


\title{Some recent developments in statistics for spatial point
  patterns}

\author{Jesper M{\o}ller and Rasmus Waagepetersen\\
Department of Mathematical Sciences, Aalborg University\\
  Fredrik Bajersvej 7G, DK-9220 Aalborg, Denmark\\email: jm@math.aau.dk, rw@math.aau.dk}


\maketitle

\begin{abstract}
This paper reviews developments in statistics for spatial point processes obtained within roughly the last decade. These developments include new classes of spatial point process models such as determinantal point processes, models incorporating both regularity and aggregation, and models where points are randomly distributed around latent geometric structures. Regarding parametric inference the main focus is on various types of estimating functions derived from so-called innovation measures. Optimality of such estimating functions is discussed as well as computational issues. Maximum likelihood inference for determinantal point processes and Bayesian inference are briefly considered  too. Concerning non-parametric inference, we consider extensions of functional summary statistics to the case of inhomogeneous point processes as well as new approaches to simulation based inference. 
\end{abstract}


\section{Introduction}

\subsection{Spatial point patterns and processes}\label{s:spps}

A spatial point pattern data set is a finite collection of points
specifying the locations of 
some `events' observed within a given spatial region 
$W$ (the `observation window'). 
Often $W$ is a $k$-dimensional compact subset of the $k$-dimensional
Euclidean space $\mathbb R^k$, 
with $k=2$ or $k=3$ in most cases, while 
other but less studied examples are
manifolds such as
$W=\mathbb S^k$, the $k-1$-dimensional sphere in $\mathbb
R^{k}$. Figures~\ref{fig:tem_bci}-\ref{fig:columns_sphere} show some examples, which are discussed in Section~\ref{s:exsdata}.


A spatial point process is a stochastic model for a spatial point
pattern data set.
To adjust for the effect of unobserved events, or if $W$ is very
large, or simply for convenience when constructing models,
the spatial point process may be defined on a possibly unbounded
region $S$ containing $W$. It may then consist of
infinitely many events.
Sometimes the events are of different types, leading to
multivariate spatial point processes. These are special
cases of marked spatial point processes, where some extra
information about each event is collected. For example, in case of the
locations of trees, each mark may specify the tree's
species or its diameter at breast height. For dimensions
$d\ge2$, there is no natural ordering on the points but sometimes an extension
to a spatio-temporal point process is considered, where the
direction of time usually 
plays a particular role. 

Today spatial point pattern analysis is widespread in many fields of
science and a rapid development of new statistical models and methods
takes place. Recent textbooks on statistics for spatial point processes include
\citet{diggle:03}, \citet{moeller:waagepetersen:03}, \citet{illian:etal:08}, and \citet{chiu:stoyan:kendall:mecke:13}. Chapter 4 of \citet{gelfand2010handbook} collects various reviews on statistics for spatial point processes. 
Moreover, \cite{baddeley:rubak:turner:15} provides a very accessible account for
applied statisticians, where the exposition is closely integrated with
analyses performed using the author's \text{spatstat} R package
\citep{baddeley:turner:05}. This package is comprehensive, flexible, and the main software for spatial point pattern analysis.

\subsection{Examples of point pattern data sets}\label{s:exsdata}

\begin{figure}
\centering
\includegraphics[width=\textwidth]{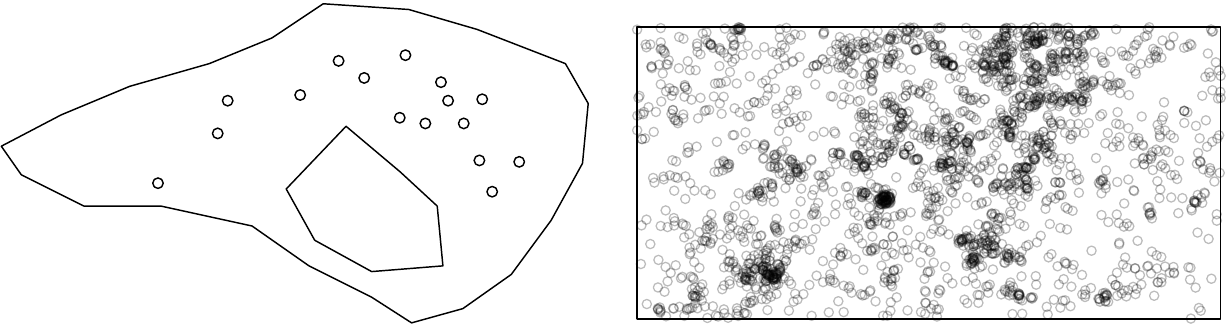}
\caption{Examples of
planar point pattern data sets. Left: locations of vesicles in a slice of a synapse from a rat. Inner polygon shows extent of a mitochondrion where  vesicles do not occur. Right: locations of Psychotria trees in the Barro Colorado Island research plot. See Section~\ref{s:exsdata} for further details.
}
  \label{fig:tem_bci}
\end{figure}

The left panel in Figure~\ref{fig:tem_bci} shows locations
of so-called vesicles in a microscopial image of a slice of a synapse
from a rat. The observation window $W$ has a non-standard form being the planar region defined by the outer
curve representing the membrane of the synapse minus the region
defined by the inner curve representing the extent of a
mitochondrion. This is a part of a large data set collected to study
whether stress affects the spatial
distribution (e.g.\ the regularity) of vesicles, see
\cite{khanmohammadi:etal:14} for further details. The vesicles data are at the microscopic scale (enclosing rectangle of $W$ has dimension 250 times 400 nm) with just 16 points that form a regular pattern, at least locally. This is in contrast to the data set in the right panel. This data set contains 2640 clustered locations of {\em Psychotria
  horizontalis} trees in the $1000 \times 500$ m Barro Colorado Island
research plot. Ecologists  study such data sets for several species to
investigate hypotheses of biodiversity, see e.g.\ \cite{hubbell:foster:83}, \cite{condit:hubbell:foster:96}, and \cite{condit:98}.

\begin{figure}
\centering
 \includegraphics[width=\textwidth]{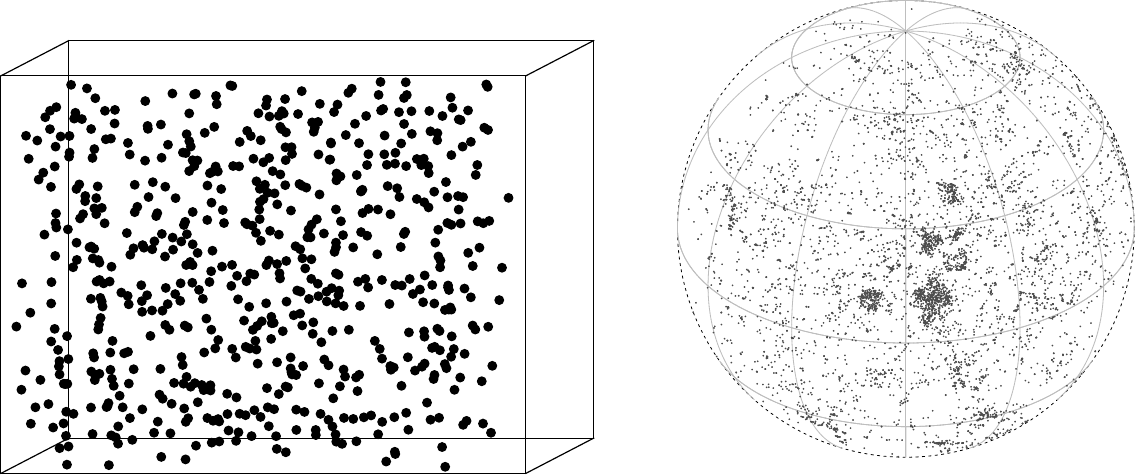} 
\caption{Examples of a point pattern data sets in the
  $3D$ space and on the sphere. Left: locations of pyramidal brain
  cells. Right: sky positions of galaxies projected onto the globe. See Section~\ref{s:exsdata} for further details.} 
\label{fig:columns_sphere}
\end{figure}

The left panel in Figure~\ref{fig:columns_sphere} shows 
the locations of 623 pyramidal cells from the Brodmann area 4 of the grey matter
of the human brain.  According  to  the  minicolumn  hypothesis  \citep{mountcastle:57}  such pyramidal
brain cells should have a columnar arrangement perpendicular to the pial
surface of the brain, and this should be highly pronounced in Brodmann area 4. However, this
hypothesis has been much debated, see \citet{AliEtAl:16}  and the references therein. It is not easy to see by eye whether there is a columnar arrangement and statistical techniques for detecting this have therefore been developed \citep{moeller:safavimanesh:rasmusssen:15}. 
The right panel shows the sky positions on the celestial sphere of 10,611 nearby galaxies catalogued in the Revised New General Catalogue and Index Catalogue, where the Milky Way obstructs the view and the point pattern is inhomogeneous and exhibits clustering, see \cite{lawrence:etal:16}.   
Statistical models and tools for point processes on the sphere have recently been developed \citep{robeson:li:huang:14,lawrence:etal:16,moeller:rubak:16}.

\subsection{Aim and outline}\label{sec:outline}

This paper is mainly
confined to a review of recent 
spatial point process models and methods for analyzing 
spatial point pattern data sets, i.e.\ without marks or times included
and with no ordering on the points. As we have a broad audience in
mind, technical details are suppressed. The paper
is to some extent a
 follow-up to our discussion paper \citet{moeller:waagepetersen:07} which gives a
concise and non-technical introduction to the theory of statistical
spatial point
pattern analysis, making analogies
with generalized linear models and random effect models. Thus we mainly focus on developments
after the publication of \citet{moeller:waagepetersen:07} 
but make no pretense of a complete literature review.

Research in statistical inference for spatial point patterns evolve around the
following three main topics:
\begin{itemize}
\item development of flexible models for point patterns exhibiting
  clustering, regularity, trends depending on spatial covariates and
  combinations of these features;
\item methodology for fitting parametric spatial point process models
  using moment relations or likelihood based methods often implemented using
  Markov chain Monte Carlo (MCMC) methods;
\item development of non-parametric summary statistics for assessing
  spatial interaction and validating fitted models.
\end{itemize}
After providing a brief introduction to basic theoretical concepts in
Section~\ref{sec:concepts}, our exposition follows the above categorization. While much research on
modelling has focused on Poisson, Cox, cluster, and Gibbs
models (briefly reviewed in Section~\ref{sec:poisson}), our modelling
Section~\ref{sec:models} 
focuses on the classes of determinantal and permanental point
processes. The potential of these models for
statistical applications has only recently been explored. In addition
models characterized by incorporating both clustering and regularity
and involving latent geometric structures are
reviewed. Section~\ref{sec:estimation} primarily considers estimating
function inference where estimating functions are obtained from
so-called innovation measures. Maximum likelihood inference for
determinantal point processes and Bayesian inference for spatial point
processes are reviewed too. Section~\ref{sec:summaries} is concerned
with functional summary statistics for non-stationary point processes
and new developments in simulation-based inference based on such
summary statistics. Finally, Section~\ref{sec:future} discusses further recent developments and perspectives for
future research.

\section{Setting and fundamental concepts}\label{sec:concepts}

This section specifies our set-up 
for spatial
point processes. For extensions
to multiple and marked point processes, and for
mathematical details, in particular measure theoretical details, the
reader may e.g.\ consult \cite{moeller:waagepetersen:03}. For instance, below $S$
and $B$ are always Borel sets and $h$ is a measurable function,
but we do not emphasize such matters.

Let $S\subseteq\mathbb R^k$ 
be the state space for some events, where typically $k=2$ or 
$k=3$ and often but not always 
$S$ is of the same dimension, cf.\ Section~\ref{s:spps}. By a
{\it spatial point process} on $S$ we mean a locally finite random subset $\bX \subset
S$. Letting $\bX_B= \bX \cap B$ for $B \subseteq S$, locally finite means that the count $N(B)=\# \bX_B$ is finite for any bounded $B\subseteq S$. 
Then the distribution of $\bX$ can be defined by two equivalent approaches:
By the finite dimensional 
distributions of the counts; or, for any  bounded $B\subseteq S$, by
specifying first the distribution of $N(B)$ and second the
distribution of the events in $\bX_B$ conditional on knowing $N(B)$.
When $S=\mathbb R^k$ and the distribution of $\bX$ is invariant under
translations in $\mathbb R^k$ respectively rotations about the origin in 
$\mathbb R^k$, 
 we say that $\bX$ is {\it stationary} respectively {\it isotropic}. 
Similarly, when $d=k-1$, $S=\mathbb S^d$ is the
$d$-dimensional sphere, and the distribution of $\bX$ is invariant under 
rotations about the origin in 
$\mathbb R^k$, we say that $\bX$ is {\it isotropic}.

For $n=1,2,\ldots$ and
pairwise disjoint bounded sets $B_1,\ldots,B_n\subseteq S$, we define
the $n$th order {\it factorial moment measure} 
$\mu^{(n)}(B)$
of the product set
$B=B_1\times\ldots\times B_n$ by the expected value of $\prod_{i=1}^n N(B_i)$.
This extends by standard methods
to a measure $\mu^{(n)}$ on $S^n$
by setting 
\[\int_{S^n} h(u_1,\ldots,u_n)\,\mathrm d \mu^{(n)}(u_1,\ldots,u_n)
=\mathrm E\sum^{\not=}_{u_1,\ldots,u_n\in\bX}h(u_1,\ldots,u_n)\]
for non-negative functions $h$ defined on $S^n$. 
Here $\not=$ over the summation sign
means that $u_1,\ldots,u_n$ are pairwise distinct. 

For simplicity and specificity, unless otherwise stated, we 
assume that $S$ is $k$-dimensional and $\mu^{(n)}$ has
a density $\rho^{(n)}$ with respect to Lebesgue measure on 
$(\mathbb R^{k})^n$ restricted to $S^n$, so that
\[\mathrm E\sum^{\not=}_{u_1,\ldots,u_n\in\bX}h(u_1,\ldots,u_n)=
\int_{S^n} h(u_1,\ldots,u_n)\rho^{(n)}(u_1,\ldots,u_n)\,
\mathrm du_1\cdots\mathrm du_n\,.\]
Then $\rho^{(n)}$ is called the $n$th order {\it joint intensity function}.
 Note that this function is uniquely defined except for a Lebesgue nullset. 
For pairwise distinct $u_1,\ldots,u_n\in S$,
we may interpret 
$\rho^{(n)}(u_1,\ldots,u_n)\,\mathrm du_1\cdots\mathrm du_n$ as the
probability of observing a point in each of $n$ infinitesimally small
volumes $\mathrm du_1,\ldots,\mathrm du_n$ containing
$u_1,\ldots,u_n$, respectively. If instead e.g.\ $S=\mathbb S^d$, then everywhere above
we should replace Lebesgue measure with $d$-dimensional surface
measure on $\mathbb S^d$. 
 
The cases $n=1,2$ are of particular interest:
$\mu=\mu^{(1)}$ is the intensity measure,
$\rho=\rho^{(1)}$ is the  {\it intensity function}, and the {\it pair
  correlation function (pcf)} is defined for $u,v\in S$ and $u\not=v$ by
\[g(u,v)=\rho^{(2)}(u,v)/\{\rho(u)\rho(v)\} \]
if $\rho(u)\rho(v)>0$, and $g(u,v)=0$ otherwise. The pcf is thus a normalized version of $\rho^{(2)}$ and it is usually easier to interpret: Roughly speaking,
the case $g(u,v)=1$ corresponds to that `$u$
and $v$ appear independently of each other', the case
$g(u,v)>1$ to that there is `attraction between $u$
and $v$' or `aggregation e.g.\ due to covariates', and the case $g(u,v)<1$ to that there is `inhibition between $u$
and $v$' or `regularity due to the model in question' (as exemplified later in (h) in Section~\ref{s:dpp}). Note that this interpretation of $g(u,v)$ may only be meaningful when $u$ and $v$ are sufficiently close. Since the diagonal in $S^2$ has zero Lebesgue measure, the definition
of $g(u,u)$ can be arbitrary. Depending on the particular model there
is often a natural choice as explained later.  

If for each $u\in S$, $\pi(u)\in[0,1]$ is a given number, then an {\it
  independent $\pi$-thinned process} $\bX_{{\mathrm{th}}}$ 
is obtained by independently
  retaining each point $u\in\bX$ with probability $\pi(u)$. It follows
  that $\bX_{{\mathrm{th}}}$ has $n$th order joint intensity function
$\rho_{{\mathrm{th}}}^{(n)}(u_1,\ldots,u_n)=
\pi(u_1)\cdots\pi(u_n)\rho^{(n)}(u_1,\ldots,u_n)$, and the pcf is the same for the two processes. 

When $S=\mathbb R^k$, {\it second order
intensity-reweighted stationarity} means that $g(u,v)=g_0(u-v)$ is
invariant under translations in $\mathbb R^k$. For example,
stationarity of $\bX$ implies second order
intensity-reweighted stationarity of both $\bX$ and 
$\bX_{{\mathrm{th}}}$. \cite{baddeley:moeller:waagepetersen:00} discuss
several other examples of second order intensity-reweighted stationary point processes.
If instead $S=\mathbb S^d$, then {\it second order
intensity-reweighted isotropy} means that $g(u,v)=g_0(d(u,v))$ is
invariant under rotations, where $d(u,v)$ is the great circle
distance. 

We sometimes refer to so-called Palm distributions. The {\it reduced Palm
distribution} of $\bX$ given a location $u \in S$ can be regarded as
the conditional distribution of $\bX \setminus \{u\}$ given that $\bX$ has a point at
$u$. We write $\bX^!_u$ for a spatial point process distributed according to
the reduced Palm distribution of $\bX$ given
$u$. In the stationary case, $\bX^!_u$ is distributed as $\bX^!_o$ translated by $u$, where $o$ denotes the origin, and so we refer to $\mathrm Eh(\bX^!_o)$ as the conditional expectation of $h$ with respect to the further points in $\bX$ given a typical point of $\bX$.  \cite{coeurjolly:moeller:waagepetersen:tutorial} give an
introduction to Palm distributions intended for a statistical audience.

A useful approach to understanding many models and methods is to consider
a subdivision of $S=\cup_{i=1}^m C_i$ into $m \ge 1$ small cells $C_i$ and  
approximate the spatial point process by a random field of presence-absence
binary variables $X_i=1[\bX \cap C_i \neq \emptyset]$ where $1[A]$ denotes indicator function of an event $A$. That is, $X_i=1$
if at least one point is present in the cell $C_i$ and $X_i=0$ otherwise. By the previously
mentioned interpretation, the joint intensities determine the
distribution of the binary variables $X_i$ associated with a subdivision of
$S$ into infinitesimally small cells $C_i$.
It is also
possible to reverse the binary random field point of view. A Poisson process
(Section~\ref{sec:poisson}) for example, can be viewed as a
limit of binary random fields of independent binary random variables
associated with a sequence of subdivisions for which the cell sizes
tend to zero. 

\section{Models}\label{sec:models}
In Section~\ref{sec:poisson}  we give a very brief
review of Poisson, cluster, Cox, and Gibbs point process models which
are covered extensively in the existing literature on spatial point
processes, cf.\ Section~\ref{s:spps}. This
is followed by more detailed reviews of
recent contributions to the modelling of spatial point processes.

\subsection{Poisson, Cox, cluster, and Gibbs point process models}\label{sec:poisson}

Assume that $\rho$ is an integrable non-negative function on
$S$ so that $\mu(B)=\int_B \rho(u) \dd u$ is finite for all bounded $B
\subseteq S$. Then $\bX$ is a {\it Poisson process} with intensity function
$\rho$ provided that $N(B)$ is Poisson distributed with mean $\mu(B)$
for any bounded $B \subseteq S$ and conditionally on $N(B)=n$, the $n$
points in $\bX_B$ are {\em iid} on $B$ with density function
proportional to $\rho$. 
The Poisson process has further independence properties:
\begin{itemize}
\item For any disjoint subsets $B_1,\ldots,B_k$, $k>1$, of $S$, 
  $\bX_{B_1},\ldots,\bX_{B_k}$ are independent.
\item A Poisson process can be viewed as a limit of binary random
  fields $\{X_i\}_{i=1}^m$ associated with increasingly fine
  subdivisions of $S$ into cells $C_i$ of volumes $|C_i|$,
  where the $X_i$ are independent with $P(X_i=1)=\rho(u_i)|C_i|$ for a
  representative point $u_i \in C_i$.
\item For any $n\ge1$, $\rho^{(n)}(u_1,\ldots,u_n)=\prod_{i=1}^n
  \rho(u_i)$.
\end{itemize}
By the latter property, $g(u,v)=1$ for any pairwise distinct $u,v \in S$ with $\rho(u)\rho(v)>0$, 
which motivates defining $g(u,u)=1$ on the diagonal of $S^2$ for a Poisson process.

Most spatial point pattern data exhibit clustering or regularity that
does not comply with the independence properties of a Poisson
process. The Poisson process is nevertheless important as a reference process for no spatial
interaction. Even more importantly, the Poisson process is the basis
for obtaining more flexible models either by specification of probability densities with respect to the Poisson
process (Section~\ref{sec:gibbs}) or by explicit constructions
 using Poisson processes as building blocks (Section~\ref{sec:coxcluster}). 

\subsubsection{Gibbs point process models}\label{sec:gibbs}

The term {\it Gibbs point processes} is in practice used as a broad term for finite point
processes specified by a density and their extensions to infinite spatial point processes obtained by considering limits for so-called local specifications, see \citet{moeller:waagepetersen:03} and the references therein. For simplicity we just consider the finite case below and review various concepts needed later. 

Denote by $\bZ$ a finite Poisson process on $S$ with intensity
function $\rho$, i.e.\ $\mu(S)<\infty$. The process $\bZ$ is used as a reference process. If $S$ is bounded, $\bZ$ is often taken to be the unit rate Poisson process with $\rho(\cdot)=1$. Denote by ${\cal N}$ the set 
of locally finite point
configurations of $S$. Suppose that $f$ is a non-negative function on ${\cal N}$ satisfying $\mathrm E f(\bZ)=1$. A spatial point process $\bX$
then has {\it density} $f$ on ${\cal N}$ (is absolute continuous) with respect to (the distribution
of) $\bZ$ provided
\begin{equation*}
\mathrm E h(\bX) = \mathrm E \left [ h(\bZ) f(\bZ) \right]
\end{equation*}
for non-negative functions $h$ on ${\cal N}$. Here, by the definition of a Poisson process,
\begin{align}
\mathrm E  h(\bZ)=\,\sum_{n=0}^\infty \frac{\exp[-\mu(S)]}{n!} 
\,\int_S\cdots\int_S h(\xn)\rho(x_1)\cdots\rho(x_n)\,\mathrm
dx_1\,\cdots\,\mathrm dx_n .\label{eq:poissonexp1}
\end{align}
In the following we consider several examples of point process densities.

If $\bX$ is a finite Poisson process with intensity function
$\af$ satisfying $\rho(u)>0$ whenever $\af(u)>0$, then $\bX$ has density
\begin{equation}\label{eq:poissonlikelihood} f(\bx)= \exp\left[\mu(S)-\int_S \af(u) \dd u \right] \prod_{u \in \bx}
\frac{\af(u)}{\rho(u)}, \quad \bx \in {\cal N},
\end{equation}
with respect to $\bZ$. If the integral involving $\af$ in \eqref{eq:poissonlikelihood}
can not be evaluated analytically, it is at least typically easy to
approximate it by numerical integration rendering maximum 
likelihood inference feasible for parametric models of Poisson processes. 

To obtain models with interaction between points, densities are
specified in terms of an 
interaction function $\phi(\cdot)\ge 0$, \jmret{
\[ f(\bx)= \prod_{\by \subseteq \bx} \phi(\by). \]
Note that $f(\bx)$ is proportional to $h(\bx)= \prod_{\by \subseteq \bx: \by \neq \emptyset} \phi(\by)$, so $\mathrm E f(\bZ)=1$ is equivalent to  
\[ \phi(\emptyset)= \frac{1}{\mathrm E h(\bZ)}= \left(\mathrm E  \prod_{\substack{\by \subseteq \bZ:\\ \by \neq
    \emptyset}} \phi(\by) \right)^{-1}. \]
Thus, $\phi(\empty)$ is the normalizing constant that upon multiplication turns $h$ into a probability density. When
specifying an interaction function it is important to check that the
expectation above defining the normalizing constant is in fact finite.} The expectation  is
typically not available in closed form and must be approximated e.g.\
by computationally intensive MCMC methods.  Often pairwise-only interaction processes are
considered in which case $\phi(\by)=1$ whenever the cardinality of
$\by$ is greater than two. The Strauss process is the well-known
example where for distinct $u,v$, $\phi(\{u\})=\phi(\{v\})=\beta>0$ and $\phi(\{u,v\})=\gm^{1[\|u-v\|
  \le R]}$ for $0<\gm\le 1$ and $R>0$. \jmret{Simulated realizations of the Strauss process for various parameter settings can be found in \citet{moeller:97aa}.}

The joint intensities of a spatial point process with density $f$ with respect
to $\bZ$ can be expressed as 
\[ \rho^{(n)}(u_1,\ldots,u_n) = \mathrm E  f(\{u_1,\ldots,u_n\} \cup \bZ) \]
for pairwise distinct $u_1,\ldots,u_n$. The expectation on the right
hand side is in general not  analytically tractable. Fast numerical approximations of the
first two joint intensities are developed in \citet{baddeley:nair:12b,baddeley:nair:12a}.

Suppose $f$ is hereditary meaning that
$f(\bx)>0$ implies $f(\by)>0$ for $\by \subseteq \bx$. Then the
$n$-point {\it conditional intensity} is defined
by
\[ \ld(u_1,\ldots,u_n,\bx) =\frac{f(\{u_1,\ldots,u_n\} \cup
  \bx)}{f( \bx)} \]
for pairwise distinct $u_1,\ldots,u_n$ and finite point configurations $\bx$
with $\bx \cap \{ u_1,\ldots,u_n\} = \emptyset$. The conditional
intensities e.g.\ play an important role in the definition of
innovation measures for Gibbs point processes, see Section~\ref{sec:estimatingfunctions}.

\subsubsection{Cox and cluster processes}\label{sec:coxcluster}

\newcommand{\bLd}{{\mathbf \Lambda}}
\newcommand{\bld}{{\mathbf \lambda}}
\newcommand{\bPhi}{{\mathbf \Phi}}

A large class of models for random spatial aggregation of points is
obtained by replacing the deterministic intensity function in a
Poisson process by a random function. More precisely, let $\mathbf
\Lambda=\{\Lambda(u)\}_{u \in S}$ be a 
non-negative random field such that  $\mathbf \Lambda$ is locally
integrable almost surely (that is, for bounded $B\subseteq S$, $\int_B\Lambda(u)\,\dd u$ exists and is
finite with probability one). Then $\bX$ is a {\it Cox process} provided that conditional on
$\bLd=\bld$, $\bX$ is a Poisson process with intensity function
$\bld$. The joint intensities are simply given by
\[ \rho^{(n)}(u_1,\ldots,u_n) = \EE \prod_{i=1}^n \Lambda(u_i)\]
for $n \ge 1$ and  pairwise distinct $u_1,\ldots,u_n \in
S$.
For a {\it log Gaussian Cox process}, i.e.\ when
$\log\Lambda$ is a Gaussian process, $\rho^{(n)}$ is just given by the Laplace transform of a multivariate normal distribution \citep{moeller:syversveen:waagepetersen:98}. 

To obtain clustered point patterns, one way is to consider
superpositions of a countable number of localized point
processes. Accordingly, a {\it Poisson cluster process} is a union
$\bX=\cup_{v \in \bPhi} \bX_{v}$ of typically finite spatial point processes
$\bX_{v}$ on $S$ indexed by a parent Poisson process $\bPhi$ on $S$ or
perhaps some other space. For convenience, conditional on $\bPhi$, the
$\bX_v$ are often assumed to be finite Poisson processes each with
intensity function $\rho_v$, in which case $\bX$ can also be viewed as
a Cox process with random intensity function $\bLd(u)=\sum_{v \in
  \bPhi}\rho_v(u)$. {\it Shot-noise Cox processes} is a specific example of
this, where $\bPhi$ is a Poisson process on $\R^d \times ]0,\infty[$ and
for a $v=(c,\gm) \in \bPhi$, $\rho_v(u)=\gm k(c,u)$, where $k(c,\cdot)$
is a probability density on $S$ \citep{moeller:03}. Simulated realizations of log Gaussian and Poisson cluster processes can be found in Chapter~5 in \cite{moeller:waagepetersen:03}.

At the expense of analytical
tractability, Poisson parent processes can be replaced e.g.\ by
regular parent processes \citep{lieshout:baddeley:02,mckeague:loizeaux:02,moeller:torrisi:05} and the Poisson distribution for cluster size could be
replaced by other integer valued distributions.
\cite{jalilian:guan:waagepetersen:13} introduce a flexible class
of shot-noise Cox processes with kernel function $k$ given by the density
of normal-variance mixture distributions. For this class of models, the pair
correlation function is given by $g(u,v)=1+c(u,v)$ where $c(u,v)$ is a
covariance function given by a scaled normal-variance mixture
density. This includes e.g.\ Mat{\'e}rn and Cauchy covariance functions.
Most examples of Cox process models are second order intensity-reweighted stationary, with an isotropic pcf. \citet{moeller:toftager:14} study Cox process models with an elliptical pcf, including shot noise Cox processes and log Gaussian Cox
processes.

The density of a finite Cox process with respect to a finite Poisson process $\bZ$
is 
\[ f(\bx)= \mathrm E f(\bx| \Ld) \]
where $f(\cdot|\bld)$ is the conditional density of $\bX$ given $\bLd=\bld$ with
respect to $\bZ$. The expectation can rarely be computed
analytically. Likelihood based inference for Cox processes therefore
typically require numerical (including Monte Carlo) approximations of the likelihood function.

\subsection{Determinantal and permanental point processes}\label{s:det-per}

\citet{macchi:75} introduced two interesting models motivated by fermions
and bosons in quantum mechanics, viz.\
determinantal and permanental point processes. 

\subsubsection{Determinantal point processes}\label{s:dpp}

{\it Determinantal point processes (DPPs)} are of interest
because of their applications in mathematical physics, combinatorics, random-matrix
theory, machine learning, and spatial statistics \citep[see][and the references therein]{LMR15}.
For DPPs on $\mathbb R^k$, 
rather flexible parametric models can be constructed and likelihood
and moment based inference procedures apply
\citep{LMR12extended,LMR15}. DPPs on the
sphere are discussed from a statistical perspective in \citet{moeller:nielsen:porcu:rubak:15} and \citet{moeller:rubak:16}.

A DPP is defined by a function $C:S\times S\mapsto\mathbb C$ as follows: $\bX$ is a
DPP with kernel $C$ if for any $n=1,2,\ldots$, $\bX$ has a non-negative $n$th
order joint intensity function given by
\begin{equation}\label{e:defdpp}
\rho^{(n)}(u_1,\ldots,u_n)=
\det\left(C(u_i,u_j)_{i,j=1,\ldots,n}\right)\quad\mbox{for all
  $u_1,\ldots,u_n\in S$}, 
\end{equation}
where $\det\left(C(u_i,u_j)_{i,j=1,\ldots,n}\right)$ 
is the determinant of the $n\times n$ matrix with 
$(i,j)$th entry $C(u_i,u_j)$. Then we write
$\bX\sim\text{DPP}(C)$. 

Before discussing the existence 
of this process, several points are in order when we assume
$\bX\sim\text{DPP}(C)$ exists (in fact it is then unique).
\begin{enumerate}
\item[(a)]
 The kernel can be
complex, since this becomes convenient when considering simulation of
DPPs as discussed below. In statistical models, it is usually real.
\item[(b)]
A Poisson process on $S$ with intensity function $\rho$ is
the special case of a DPP where $C(u,u)=1$ and 
$C(u,v)=0$ for $u\not=v$. 
\item[(c)]
The intensity
function is $\rho(u)=C(u,u)$, and the pcf is 
$g(u,v)=1-|C(u,v)|^2/\{C(u,u)C(v,v)\}$. Thus $g(u,u)=0$. 
\item [(d)]
By \eqref{e:defdpp}, an independent $\pi$-thinning of 
$\bX\sim\text{DPP}(C)$ results in 
$\bX_{{\mathrm{th}}}\sim\text{DPP}(C_{{\mathrm{th}}})$, where
$C_{{\mathrm{th}}}(u,v)=\sqrt{\pi(u)\pi(v)}C(u,v)$. Thus, if
$C(u,v)=C_0(u-v)$ is stationary, $\bX_{{\mathrm{th}}}$ is second order
intensity-reweighted stationary (on the sphere this holds provided $C(u,v)=C_0(d(u,v))$ is isotropic).
\item[(e)]
In particular, for
any $B\subseteq S$, $\bX_B$ is a DPP on $B$, with the
restriction of $C$ to $B\times B$ as its kernel. This is in contrast
to Gibbs point processes. For example, the restriction of a Strauss process to a
smaller region is not a Strauss process and its density is intractable even up to a constant of proportionality. 
\item[(f)]
A smooth one-to-one transformation of
$\bX$ results in a new DPP, cf.\ \citet{LMR12extended}.
\item[(g)]
By \eqref{e:defdpp} and since $\rho^{(n)}$ is
non-negative, $C$ has to be positive semi-definite. In fact,
in most work on DPPs, the kernel is assumed to be Hermitian, i.e.\  
$C$ is
a complex covariance
function. In the sequel,
we make this assumption.  
\item[(h)]
\jmret{By (c)} this implies that $g\le1$, which usually is a strict inequality,
i.e.\ there is `regularity at any scale'. In fact, an even stronger
result holds: for all $u_1,\ldots,u_n\in S$ we have
$\rho^{(n)}(u_1,\ldots,u_n)\le\rho(u_1)\cdots\rho(u_n)$, with equality
if and only if $\bX$ is a Poisson process with intensity function
$\rho$. 
\end{enumerate}

The existence of $\bX\sim\text{DPP}(C)$
relies on a spectral representation of the kernel: Consider
$C$ restricted to any compact
set $B\subseteq S$. Under mild conditions \citep[e.g.\ continuity of $C$ is
sufficient, cf.][]{LMR12extended,LMR15}, for any $u,v\in B$,
\begin{equation}\label{e:spec}
C(u,v)=\sum_{i=1}^\infty\lambda_i\phi_i(u){\overline{\phi_i(v)}},
\end{equation}
where $\overline z$ denotes complex conjugate of $z\in\mathbb C$,  $\lambda_i$ is the
eigenvalue corresponding to the eigenfunction $\phi_i$, and $\int_B \phi_i(u){\overline{\phi_j(u)}}\,\mathrm du=1[i=j]$.  
Then existence is equivalent to that all $\lambda_i\le1$, cf.\ \citet{LMR12extended,LMR15} and the references therein.
If $S=\mathbb R^k$ and $C(u,v)=C_0(u-v)$ is stationary, under mild conditions existence is  equivalent to that the
spectral density for $C_0$ is at most 1 \citep{LMR12extended,LMR15}. 
If
$S=\mathbb S^{d}$ and $\bX$ is isotropic, 
 the eigenfunctions are given by
spherical harmonics, see \citet{moeller:nielsen:porcu:rubak:15}. 

\citet{LMR12extended,LMR15} conclude that 
DPPs offer relatively flexible models
for repulsiveness, although less flexible than Gibbs point processes, \jmret{e.g.\ DPPs cannot be as repulsive as Gibbs hard-core point processes can be}. 
This is due to 
the bound on the spectrum of $C$ needed for the existence of 
$\bX\sim\text{DPP}(C)$: The bound
implies a trade-off between a large intensity and a strong degree of
interaction. 

The spectral decomposition
allows a closed form expression for the likelihood when $\bX$ is
observed within a compact region, but in practice the likelihood needs to be approximated as discussed later 
 in 
Section~\ref{s:mle}. Furthermore, $N(B)$ is distributed as
$\sum_{i=1}^\infty N_i$, where $N_1,N_2,\ldots$ are independent
Bernoulli variables with parameters
$\lambda_1,\lambda_2,\ldots$. Simulation of $N_1,N_2,\ldots$ is easy 
and conditional on $N_1,N_2,\ldots$, the
events in $\bX_B$ can be generated one by one 
using a fast exact simulation procedure based on the fact that $\bX_B$
has a density which is proportional to the $(\sum_{i=1}^\infty N_i)$th order intensity
function for a DPP with kernel 
$\sum_{i=1}^\infty N_i\phi_i(u){\overline{\phi_i(v)}}$. See 
\citet{Hough:etal:06} and \citet{LMR12extended,LMR15}. 

To summarize, in comparison to  
Gibbs point processes, DPPs possess many appealing properties: DPPs can be easily simulated; their moments 
and the likelihood are
tractable; a DPP restricted to a smaller region is
again a DPP; an independent thinning of a DPP or a smooth one-to-one
transformation is again a DPP; and the
distribution of the number of events within a compact region is 
known. 

\subsubsection{Permanental point processes and extensions}\label{sec:perm}

Given a complex covariance
function $C:S\times S\mapsto\mathbb C$, $\bX$ is a
 {\it permanental point process 
(PPP)} with kernel $C$ if for any $n=1,2,\ldots$, $\bX$ has a non-negative $n$th
order intensity function as in \eqref{e:defdpp} but with the
determinant replaced by the permanent
\[\per\left(C(u_i,u_j)_{i,j=1,\ldots,n}\right)=\sum_{(\sigma_1,\ldots,\sigma_n)}\prod_{i=1}^n
C(u_{\sigma_i},u_{\sigma_j}),\]
where the sum is over all permutations of $1,\ldots,n$. In fact,  
$\bX$ is then a Cox process driven by $|\bZ|^2$, where $\bZ$
is a 
zero-mean complex Gaussian process 
with covariance function $C$. 
The intensity function is $\rho(u)=C(u,u)$ and the pcf is $g(u,v)=1+|C(u,v)|^2/\{C(u,u)C(v,v)\}$.
This implies that $g\ge1$ which usually is a strict inequality,
i.e.\ there is `aggregation at any scale'.

Despite some attractive properties of PPPs \citep{Hough:etal:06,mccullagh:moeller:05}, including 
that the spectral decomposition \eqref{e:spec}
allows a closed form expression for the likelihood when $\bX$ is
restricted to a compact set, PPPs
have yet not been used much in applications, probably because computing 
permanents is computationally infeasible even for relatively small
matrices. Since $1\le g\le2$, PPPs may be a rather restrictive class
of models for aggregation. Nevertheless, we think PPPs deserve to be
investigated more because of their attractive moment properties.  

Determinantal and permanental point processes can be extended
by replacing the determinant or permanent with a weighted determinant
or permanent, see \citet{mccullagh:moeller:05}. This extension includes the case of 
a Cox process driven by $\sum_{i=1}^k\bY_i^2$, where $\bY_1,\bY_2,\ldots$
are independent zero-mean real Gaussian processes with a common real 
covariance function $C$.

\subsection{Regularity on the small scale and aggregation on the large scale}

In the classical spatial point process literature, spatial point processes are often classified into
three main cases, viz.\ complete spatial randomness (i.e.\ the Poisson process),
regularity (e.g.\ DPPs and most Gibbs point
processes),
and aggregation (e.g.\ Cox processes). This can be too simplistic, and
often we need a model with aggregation on the large scale and
regularity on the small scale. 

Gibbs point processes with an inhomogeneous first order potential and 
inhibitive pairwise interactions are models for deterministic
aggregation combined with inhibition at the small scale.  Alternatively, in order to introduce inhomogeneity,  
homogeneous regular Gibbs processes can be subjected to location dependent thinning, smooth transformation, or local scaling 
\citep{baddeley:moeller:waagepetersen:00,jensen:nielsen:00,hahn2003}. Another pos\-si\-bi\-li\-ty is to consider a homogeneous Gibbs
point process with a well-chosen higher-order potential that incorporates
inhibition at small scales and attraction at large scales. A famous
example is the Lennard-Jones pair-potential
\citep{ruelle:69}. Another specific potential of this type can be
found in \cite{goldstein2014}. Disadvantages of Gibbs point process models or models derived from those are that densities and joint intensities are intractable and that simulation requires elaborate MCMC methods, cf.\ Section~\ref{sec:gibbs}. 

In the following we focus on two modeling strategies  to obtain {\it random aggregation as well as random local regularity}. In \citet{andersen:hahn:15} the starting point is a (aggregated) Cox process which is subjected to dependent thinning to create local regularity. Conversely, \cite{lavancier:moeller:16}
 begin with a (regular) determinantal point process and use randomly spatially varying thinning to obtain aggregation. 
 
Specifically, 
 \citet{andersen:hahn:15} apply Mat{\'e}rn type II dependent thinning to shot-noise Cox processes (Section~\ref{sec:coxcluster}). For a given point pattern and a specified distance $h$, Mat{\'e}rn II thinning acts by first attaching random positive marks (`arrival times') to each point. Subsequently a point is removed if it has a neighbour within distance $h$ and with a smaller mark (i.e.\ the neighbour arrived earlier). For a given location $u$, \citet{andersen:hahn:15} define the retention probability at $u$ as the ratio between the intensities of the thinned process and the orginal process at $u$. They derive expressions for the retention probabilities in terms of integrals that, however, must be evaluated using numerical integration. Simple approximate formulae for the intensity function and the second-order joint intensity of the thinned process are also provided and used for fitting of parametric models to data sets of cell centres in microscopial images of bone marrow tissue. 

On the other hand, \citet{lavancier:moeller:16} consider a spatial point process $\bY$ on  $S$ and a random field $\Pi=\{\Pi(x):x\in S\}$ of retention probabilities independent of $\bY$. Conditional on $\Pi$, $\bX$ is then an independent thinning of $\bY$ with retention probabilities $0 \le \Pi(u) \le 1$ at $u \in S$. In
\citet{stoyan:79} and \citet{chiu:stoyan:kendall:mecke:13},
 $\bX$ is called an interrupted
point process (which is a good terminology when each $\Pi(x)$ is
either 0 or 1). Simple relations between the intensity and pair correlation functions of $\bY$ and $\bX$ are then available:
\begin{equation}\label{e:intrel}
\rho_{\bX}(u)=q(u) \rho_{\bY}(u),\quad q(u)=\mathrm E\Pi(u),\quad u\in S,
\end{equation} 
and (setting $0/0=0$)
\begin{equation}\label{e:pcfs}
g_{\bX}(u,v)=M(u,v)g_{\bY}(u,v),\quad
M(u,v)=\frac{\mathrm E[\Pi(u)\Pi(v)]}
{\mathrm E[\Pi(u)]\mathrm E[\Pi(v)]},\quad u,v\in S.
\end{equation} 
For example, if $g_{\bY}\le 1$ while $\Pi$ is positively correlated (i.e.\ $M>1$), then we may achieve
$g_{\bX}$  smaller/larger than
1 on the small/large scale. In the cases where $\bY$ is a determinantal point process or a Mat{\'e}rn hard core model of type I or II 
\citep{matern:86} and $\Pi$ is based on a transformed Gaussian process
or the characteristic function of a Boolean model of balls,  \citet{lavancier:moeller:16} derive explicit formulae for  $\rho_\bX$ and $g_{\bX}$. For such models, simulation of $(\bY,\Pi,\bX)$ restricted to a bounded region is furthermore straightforward. Conditional simulation of $\Pi$ given $\bX$, $\bY$ or both is more complicated and discussed in \citet{lavancier:moeller:16}.

\subsection{Latent geometric structures}\label{s:latent}

Spatial point processes in the neighbourhood of a reference
structure are frequently observed. Often the reference structure has a kind of linear structure. A diverse set of examples at very different
scales are
 gold coins near Roman roads 
\citep[pp.~226--229 in][]{hodder:orton:76}, 
copper deposits in the neighbourhood of lineaments
 \citep{berman:86,baddeley:turner:06,illian:etal:08},
 galaxies at the boundary of
cosmic voids \citep{icke:weygaert:87,weygaert:icke:89,weygaert:94},
  pores at the boundary of grains
\citep{karlsson:liljeborg:94,skare:moeller:jensen:06}, 
animal latrines near territorial
boundaries \citep{blackwell:01,blackwell:moeller:02,skare:moeller:jensen:06}, 
linear rows of mines \citep{walsh:raftery:02}, 
specific tree species along rivers in a rain forest
\citep{valencia:etal:04},
and 
brain
cells with a columnar structure (cf.\ Figure~\ref{fig:columns_sphere}). In many cases
the reference structure is not easy to recognize. The objective
of the statistical analysis of point patterns of this type
may either be to describe the distribution of the point pattern or
to reconstruct the reference structure or perhaps both.

For example, as in
\citet{blackwell:01}, \cite{blackwell:moeller:02}, and \citet{skare:moeller:jensen:06}, the
unknown reference structure may be modelled by a Voronoi tessellation
and the points of an unknown point process on the edges of this
tessellation are randomly disturbed to produce an observed point
pattern with points around the edges. Such complex models are analyzed
in a Bayesian MCMC setting with priors on the spatial point process of nuclei
generating the Voronoi tessellation, the point process on the edges,
and the model parameters.
Typically, for tractability, the spatial point process priors are Poisson
processes, in which case the likelihood for the observed point pattern
is described by a Cox process where the driving intensity is specified
by the nuclei and the model parameters. The posterior then just concerns the Voronoi tessellation and the model parameters, and
a major element of the MCMC algorithm is the reconstruction of the Voronoi tessellation after a proposed local change of the tessellation.  

The construction above is in \citet{moller:rasmussen:15} extended to construct models for cluster point processes within territories modelled by the Voronoi cells, where conditional on the territories/cells, the clusters are independent Poisson processes whose points may be aggregated around or away from the nuclei and along or away from the boundaries of the cells. 

To model a columnar structure for the pyramidal cell data set in Figure~\ref{fig:columns_sphere}, a hierarchical construction starting with a Poisson line process $\mathbf L$ is used in
\citet{moeller:safavimanesh:rasmusssen:15}: Consider on each line $l_i$ of $\mathbf L$ a Poisson process $\bY_i$ and let $\bX_i$ be obtained by random displacements of the points in $\bY_i$. Then $\bX$ is the superposition of all the $\bX_i$. For this model \citet{moeller:safavimanesh:rasmusssen:15} discuss moment results and simulation procedures based on the fact that $\bX$ becomes a Cox process which is closely related to a shot-noise Cox process where the centre process is replaced by $\mathbf L$. 
Simulation of $\bX$ within a bounded observation window needs to take care of edge effects, and conditional simulation of its random intensity (when $\bX$ is viewed as a Cox process) requires MCMC techniques; for details, see \citet{moeller:safavimanesh:rasmusssen:15}. 
In the case where all the lines are parallel and their direction is known (this is a reasonable assumption for the pyramidal brain cell data set), the pcf is tractable and useful for inference. 

\section{Estimation}\label{sec:estimation}

In this section we assume that $\bX$ is observed within a bounded set $W\subseteq S$ and that a parametric model has been specified for
the distribution of $\bX$ or alternatively just for
the joint intensities $\rho^{(n)}$ for some $n \ge 1$ (often just the intensity and pair correlation functions are considered, i.e.\ $n=1,2$). 
The unknown parameter to be
estimated is denoted $\ta$ and assumed to be real of dimension $p$, and we 
 write $\rho^{(n)}(\cdot;\ta)$, $\ld^{(n)}(\cdot;\ta)$
etc.\ to stress the dependence on $\ta$. 

In the case of a Gibbs process where the likelihood is specified, maximum likelihood inference and
  Bayesian inference is
  in general
difficult due to the complicated normalizing constant. The
Poisson process is one exception where the modest computational
challenge is the evaluation of an integral over the observation window
$W,$ cf.\ \eqref{eq:poissonlikelihood}. For Cox and cluster processes,
the likelihood function is also very complicated, involving expectations
with respect to the unobserved random intensity function or the
unobserved parent points. Approximate maximum likelihood
or Bayesian inference may be implemented using Monte Carlo methods or
Laplace approximations \citep[see e.g.][for
reviews]{moeller:waagepetersen:03,moeller:waagepetersen:07}. Some comments on maximum likelihood inference
for determinantal point processes are given in Section~\ref{s:mle}
while Section~\ref{sec:bayesinference} discusses some points regarding
Bayesian inference.

Due to the computational obstacles related to likelihood based inference,
much interest has focused on establishing computationally easier
approaches for spatial point processes with specified conditional intensity in the case of Gibbs processes and specified
intensity and pair
correlation functions in the case of Cox and cluster processes. This includes Takacs-Fiksel and pseudolikelihood estimation for
Gibbs point processes
\citep[e.g.][]{besag:77,fiksel:84,takacs:86,jensen:moeller:91,diggle:etal:94,billiot:97,baddeley:turner:00,billiot:coeurjolly:drouilhet:08,coeurjolly:dereudre:lavancier:12}
and various types of minimum contrast, composite
likelihood, Palm likelihood and estimating functions for Cox and cluster processes and DPPs
\citep[e.g.][]{schoenberg:05,guan:06,waagepetersen:07,tanaka:ogata:stoyan:08,waagepetersen:guan:09,dvorak:prokesova:12,prokesova:jensen:13,prokesova:dvorak:jensen:15,guan:jalilian:waagepetersen:15,zhuang:15,LMR12extended,LMR15}. As discussed in Sections~\ref{s:4.1}-\ref{s:4.3}, all of these methods belong to a common framework of estimating
functions based on signed innovation measures \citep{baddeley:etal:05,waagepetersen:05,zhuang:06}.

\subsection{Innovation measures and estimating functions}\label{sec:estimatingfunctions}\label{s:4.1}

 For a spatial point process with
$n$'th order joint intensity 
$\rho^{(n)}$, we define the $n$'th order {\it innovation
measure} as
\begin{align*} & I^{(n)}(B_1  \times \cdots \times B_n)\\  = & 
\sum_{u_1,\ldots,u_n \in \bfX}^{\neq} 1[u_1 \in B_1, \ldots, u_n \in
B_n]- \int_{B_1 \times \cdots \times B_n}
\rho^{(n)}(u_1,\ldots,u_n)\, \dd u_1 \cdots \dd u_n 
\end{align*}
for $B_i \subseteq S$, $i=1,\ldots,n$.
By definition of the factorial moment measure, $I^{(n)}(B_1 
\times \cdots \times B_n)$ has expectation zero. 
For Gibbs point processes the factorial moment measure is
not known in closed form and it is more convenient to define
{\it conditional innovation measures} in terms of the $n$'th order conditional
intensity: For $F$ a set of locally finite point configurations and
$B_i$ as above,
\begin{align*}  &I^{(n)}(B_1 \times \cdots \times B_n \times
  F|\bX) \\  = &
\sum_{u_1,\ldots,u_n \in \bX}^{\neq} 1[u_1 \in B_1, \ldots ,u_n \in
B_n, \bX \setminus \{ u_1,\ldots,u_n\} \in F] \\ & -  \int_{B_1\times\cdots\times B_n}  1[\bX \in F]
\ld^{(n)}(u_1,\ldots,u_n,\bX) \,\dd u_1 \cdots \dd u_n  \end{align*}
which has expectation zero by the Georgii-Nguyen-Zessin
formula \citep{georgii:76,nguyen:zessin:79b}.

Introducing the dependence on $\ta$ as well as $p$-dimensional
weight functions $h(\cdot;\ta)$ on $S^n$ or on $S^n \times {\mathcal N}$, we then obtain
classes of $n$'th order {\it unbiased estimating functions}
\begin{align} e_{h}^{(n)}(\ta)  = & \int_{W^n} h(u_1,\ldots,u_n;\ta)
  I^{(n)}(\dd u_1\cdots  \dd u_n;\ta) \nonumber
\\ = & \sum_{u_1,\ldots,u_n \in \bX_W}^{\neq} h(u_1,\ldots,u_n;\ta)-
\int_{W^n}h(u_1,\ldots,u_n;\ta) \rho^{(n)}(u_1,\ldots,u_n;\ta)\, \dd u_1 \cdots
\dd u_n  \label{eq:erhon}
\end{align}
or
\begin{align} e_{h}^{\text{tf},(n)}(\ta) = & \int_{W^n \times {\mathcal N}} h(u_1,\ldots,u_n,\bz;\ta)
I^{(n)}(\dd u_1\cdots \dd u_n\dd \bz |\bX;\ta) \nonumber \\
= & \sum_{u_1,\ldots,u_n \in \bX_W}^{\neq} h(u_1,\ldots,u_n,\bX
\setminus \{u_1,\ldots,u_n\};\ta) \nonumber \\ & -
\int_{W^n}h(u_1,\ldots,u_n,\bX;\ta) \lambda^{(n)}(u_1,\ldots,u_n,\bX;\ta)\, \dd u_1 \cdots
\dd u_n .\label{eq:etf}
\end{align}
Here unbiasedness means that $\mathrm E e_{h}^{(n)}(\ta)=\mathrm E e_{h}^{\text{tf},(n)}(\ta) =0$, with the expectations calculated under $\theta$.
We will use the term {\it Takacs-Fiksel estimating function} for an estimating
function of the type \eqref{eq:etf}. For Takacs-Fiksel estimating functions we may account for edge effects by letting $W$ be an eroded observation window, see e.g.\ \citet{moeller:waagepetersen:03}.

\subsection{Estimating functions based on joint intensity functions}\label{sec:est_intensities}

In practice estimating functions of the form \eqref{eq:erhon} are
mainly considered in the first ($n=1$) or second order ($n=2$)
cases and can be motivated by composite likelihood arguments.

\subsubsection{Composite likelihood}\label{s:compositelikelihood}
The first notion of {\it composite likelihood}  for spatial
point processes is due to \cite{guan:06} who defined a
bivariate probability density $f(u,v;\ta) \propto \rho^{(2)}(u,v;\ta)$,
$u,v \in W,$ and considered the log composite likelihood $\sum_{u,v \in
  \bX}^{\neq}\log f(u,v;\ta)$. Below we discuss and compare other approaches to composite likelihood for spatial point processes based on
the infinitesimal interpretation of joint intensity functions.



Letting $h(u;\ta)=  \dd\log \rho(u;\ta)/\dd\ta$ in
\eqref{eq:erhon} in the case $n=1$, the score of the Poisson log
likelihood is recovered. If the underlying spatial point process is not
Poisson, the estimating function can be interpreted as a composite
likelihood score for the binary indicators $X_i$ of presence of points in cells
of an infinitesimal partition of the observation window mentioned in
the end of Section~\ref{sec:concepts} \citep{schoenberg:05,waagepetersen:07,guan:loh:07,moeller:waagepetersen:07}.
Likewise, in case $n=2$ with $h(u,v;\ta)= 1[\|u-v\| \le R]\,\dd \log
\rho^{(2)}(u,v;\ta)/\dd \ta$ for some tuning parameter $R>0$, the score of a second order composite likelihood

\begin{equation}\label{eq:secondorder}
\sum_{u,v \in \bX_W}^{\neq} 1[\|u-v\| \le R] \frac{\dd}{\dd \ta} \log
\rho^{(2)}(u,v;\ta) - \int_{W^2} 1[\|u-v\| \le R] \frac{\dd}{\dd \ta} \rho^{(2)}(u,v;\ta) \dd u \dd v
\end{equation}
is obtained --- this time for binary indicators $X_i X_j$ of simultaneous
occurrence of points in $R$-close pairs of distinct cells $C_i$ and $C_j$  of the aforementioned
partition \citep{waagepetersen:07,moeller:waagepetersen:07}. 
Returning
to \cite{guan:06}'s composite likelihood the associated score
is
\begin{equation}\label{eq:cl2score} \sum_{u,v \in \bX_W}^{\neq} 1[\|u-v\| \le R] \frac{\dd}{\dd \ta} \log
\rho^{(2)}(u,v;\ta) -   \sum_{u,v \in \bX_W}^{\neq}
\frac{ \int_{W^2} \frac{\dd }{\dd \ta} 1[\|u-v\| \le
  R]\rho^{(2)}(u,v;\ta) \,\dd u \dd v}{ \int_{W^2} 1[\|u-v\| \le R] \rho^{(2)}(u,v;\ta) \,\dd u
  \dd v}. 
\end{equation} 
The estimating functions \eqref{eq:secondorder} and \eqref{eq:cl2score} are closely related since their first terms agree. Moreover, the expectation of the last term in \eqref{eq:cl2score} coincides with the last term in \eqref{eq:secondorder} so that \eqref{eq:cl2score} is also unbiased.

The so-called {\it Palm likelihood} \citep{tanaka:ogata:stoyan:08} is based on composite likelihood
arguments too. For a fixed point $u \in \R^d$ the reduced Palm
spatial point process $\bX_u^!$ has intensity function $\rho(v|u;\ta)=\rho^{(2)}(u,v;\ta)/\rho(u;\ta)$. Assuming that $\bX$ is stationary with known constant intensity, the unbiased 
score \citep{prokesova:jensen:13} of
the Palm likelihood is 
\begin{equation}\label{eq:palmlike} \sum_{u \in \bX_W} \sum_{v \in \bX \setminus u} \frac{\dd}{\dd\theta}
\log \rho(v|u;\ta)1[\| v-u \| \le R] - N\left(\bX_W\right)\int_{\|v\|\le R} 
\frac{\dd }{\dd \ta} \rho(v |o) \dd v.
\end{equation}
 Obviously, considering the first term in \eqref{eq:palmlike},
the Palm likelihood is closely related to the two other types of second order
composite likelihoods. 

Asymptotic results for the various types of composite likelihoods are
provided in
\cite{guan:06}, \cite{waagepetersen:07}, \cite{guan:loh:07}, \cite{waagepetersen:guan:09}, and \cite{prokesova:jensen:13}. 
It is not known which type of composite likelihood is
most efficient. In any case, none of them are optimal, see Section~\ref{sec:ql}.

\subsubsection{Quasi-likelihood}\label{sec:ql}

While the estimating functions described in the previous setting have
a nice motivation in terms of composite likelihoods, they are in
general not optimal. Optimality can be achieved following the route of
quasi-likelihood. 

For an $m \times 1$ data vector $Y$ with mean vector $\mu$
depending on a $p$-dimensional parameter $\ta$, a class of unbiased estimating functions
are given by
\[ A^\T(Y -\mu) \]
for $m \times p$ matrices $A$. This can be viewed as a
matrix-vector analogue of the estimating function \eqref{eq:erhon} in
case $n=1$ with
$Y-\mu$ and $A$ corresponding to the innovation process $I^{(n)}$ and
the weight function
$h$, respectively. The optimal choice of $A$ is the solution of $V A =D$ where $D$ is
the $m\times p$ matrix of partial derivatives $\dd \mu_i/\dd \ta_j$
and $V$ is the covariance matrix of $Y$. This choice of $A$ yields the
so-called quasi-likelihood score
\citep[e.g.][]{heyde:97}. 

\cite{guan:jalilian:waagepetersen:15}
generalize the concept of {\it quasi-likelihood} to the case of spatial
point processes. Considering the class of first order estimating
functions \eqref{eq:erhon} they identify the optimal weight function $h$ as the solution of a Fredholm integral
equation 
\[h + T h = \frac{\dd}{\dd \ta} \log \rho(\cdot,\ta) \]
where $T$ is a certain integral operator depending on the pair
correlation function of the spatial point process. 
They further establish asymptotic
properties of the spatial point process quasi-likelihood and also
provide an efficient numerical implementation of the method. 

\cite{guan:etal:16} take the quasi-likelihood idea further and consider a second order quasi-likelihood giving
the optimal choices of $h_1$ and $h_2$ for the linear combination
\[ e^{(1)+(2)}_{h_1,h_2}(\ta)= e^{(1)}_{h_1}(\ta)+
e^{(2)}_{h_2}(\ta). \]
This leads to considerable computational challenges as the optimal
functions $h_1$ and $h_2$ now depend also on third and fourth order
joint densities. However, in the stationary case, \cite{guan:etal:16} show that
restricting attention to constant functions $h_1$ and functions $h_2$
with $h_2(u,v)$ only depending on $v-u$ does not lead to a loss of
efficiency asymptotically and moreover simplifies computations
greatly.

Given a set of estimating functions
$e_1(\ta),\ldots,e_m(\ta)$, 
a related topic is to obtain a combined estimating function as an optimal
linear combination $e(\ta)= \sum_{i=1}^m A_i e_i(\ta)$ for $p \times p$ matrices $A_i$. The
weight matrices $A_i$ can be determined according to quasi-likelihood arguments and
this approach is considered in \cite{deng:waagepetersen:guan:14} for
estimation of parameters in stationary point processes. Similarly,
\cite{lavancier:rochet:16} consider how to obtain an estimate as an optimal linear
combination $\hat \ta= \sum_{i=1} w_i \hat \ta_i$ given a collection
of estimates $\hat \ta_1,\ldots,\hat \ta_m$ of the same parameter $\ta$.

\subsubsection{Case-control  likelihood}\label{sec:logisticlikelihood}

Consider \eqref{eq:erhon} with $n=1$ and $h(u)= \dd\log \rho(u;\ta)/\dd \ta$.
Suppose in addition to $\bX$ another spatial point process $\bZ$ of intensity
$\af$ is available where $\bX$ and $\bZ$ are almost surely disjoint. Then 
\begin{equation}\label{eq:integralapprox} \sum_{u \in \bX \cup \bZ} \frac{\dd \rho(u;\ta)/\dd
  \ta}{\rho(u;\ta)+\af(u)} 
\end{equation}
is an unbiased estimate of the last integral in
\eqref{eq:erhon}. The estimating function obtained by replacing the
integral by the unbiased estimate \eqref{eq:integralapprox} is the derivative of 
\begin{equation}\label{eq:logisticlikelihood} \sum_{u \in \bX} \log \frac{\rho(u;\ta)}{\rho(u;\ta)+\af(u)} + \sum_{u
  \in \bZ} \log \frac{\af(u)}{\rho(u;\ta)+\af(u)} 
\end{equation}
which is the limit of log conditional likelihoods based on the conditional
distribution of binary indicators $X_i$ given $X_i+Z_i=1$ where the
$Z_i$ are presence/absence indicators for $\bZ$. Thus, in
epidemiological terminology, $\bX$ and $\bZ$ play the role of case and
control processes. 

\cite{diggle:rowlingson:94} introduced
\eqref{eq:logisticlikelihood} in  the case
where $\bZ$ is a spatial point process representing a background population
and $\bX$ is a spatial point process of disease cases with intensity function
\[ \rho(u;\ta)=f(u;\ta) \af(u). \]
In this setup $\af$ is assumed to be unknown but cancels out in both
terms of \eqref{eq:logisticlikelihood}. \cite{waagepetersen:08}
considers the case where $\bZ$ is a user generated dummy point
process generated with the sole purpose of approximating the
integral. \cite{guan:waagepetersen:beale:08} generalize
\eqref{eq:logisticlikelihood} to the case of second order joint
intensities. \cite{diggle:etal:10}, \cite{huang:etal:14}, and \cite{chang:etal:15} further exploit and expand the case control
methodology for spatial point processes in
epidemiological applications with multiple sources of data.

\subsection{Pseudo-likelihood}\label{s:4.3}

Estimating functions of the type \eqref{eq:etf} have so far mainly been
considered when $n=1$ in which case the by far most popular
weight function is $h(u,\bx)=\dd \log \ld(u,\bx;\ta) / \dd \ta$
leading to the pseudo-likelihood score \citep[e.g.][]{besag:77,jensen:moeller:91,baddeley:turner:00,billiot:coeurjolly:drouilhet:08}. A review of
other choices of weight functions is provided in \cite{coeurjolly:dereudre:lavancier:12}
who also introduce weight functions designed to lead to
computationally quick Takacs-Fiksel estimating functions in case
of Strauss and quermass point processes
\citep{kendall:lieshout:baddeley:99}. In this section we focus on the {\it pseudo-likelihood}
approach.

As for the Poisson likelihood the computational issue with the 
pseudo-likelihood score is the evaluation of the integral in
\eqref{eq:etf}. \cite{baddeley:turner:00} suggested a
computationally efficient numerical quadrature approximation based on the
\cite{berman:turner:92} device. Formally, the approximated pseudo-likelihood takes the form of
a Poisson regression so that it can be easily implemented using standard statistical
software for generalized linear models. One problem with the
Berman-Turner approach is that the approximate pseudo-likelihood score
is not unbiased which can lead to a strongly biased estimate unless a large number of
quadrature points is used. 

As an alternative to the Berman-Turner device, \cite{baddeley:etal:14} proposed
an unbiased approximation of the pseudo-likelihood score following the
case-control approach in Section~\ref{sec:logisticlikelihood} with
$\rho(u;\ta)$ replaced by $\ld(u,\bX)$ in
\eqref{eq:logisticlikelihood}. The approximated unbiased score is 
\begin{equation}\label{eq:logisticregression} \sum_{u \in \bX} \frac{(\dd \log \ld(u,\bX;\ta) / \dd \ta)
  \af(u)}{\ld(u,\bX;\ta)+\af(u)} - \sum_{u \in \bZ} \frac{(\dd \log \ld(u,\bX;\ta) / \dd \ta)}{\ld(u,\bX;\ta)+\af(u)}
\end{equation}
where $\bZ$ is a user generated spatial point process of random quadrature
points independent of $\bX$ and with intensity function $\af$. 
In the common case where $\ld(u,\bX;\theta)$ is of log-linear form,
\eqref{eq:logisticregression} is formally equivalent to the score of a
logistic regression with probabilities of the form $\ld(u,\bX;\ta)/\{
\ld(u,\bX;\ta) + \af(u)\}$ where $-\log \af(u)$ plays the role of a
known offset. Thus also \eqref{eq:logisticregression} is very
easily implemented using standard software for generalized linear
models and is available in \texttt{spatstat}. \cite{baddeley:etal:14}
establish asymptotic consistency and normality of estimates obtained
using \eqref{eq:logisticregression}. The use of random quadrature
points $\bZ$ adds additional estimation error compared with the
exact pseudo-likelihood. However, in terms of mean square error,
\eqref{eq:logisticregression} outperforms the Berman-Turner
implementation of the pseudo-likelihood due to less bias.


\subsection{Maximum likelihood for DPPs}\label{s:mle}

We now discuss maximum likelihood inference for a determinantal point process $\bX\sim\mathrm{DPP}(C)$ restricted to a compact space $S$ and based on a realization 
$\bX=\{\bx_1,\ldots,\bx_n\}\subset S$. We can make this assumption without loss of generality, cf.\ (e) in Section~\ref{s:dpp}. 
We assume that the eigenvalues of $C$ as given in \eqref{e:spec} satisfies $\lambda_i<1$, $i=1,2,\ldots$. 
Then $\bX$ has a density 
\[f(\bx)=\exp(|S|-D) \,\text{det}\left(\tilde C(x_i,x_j)_{i,j=1,\ldots,n}\right)
\]
with respect to the unit rate Poisson process on $S$, where 
\[D=-\log\mathrm P(\bX=\emptyset)=-\log\sum_{i=1}^\infty\log(1-\lambda_i)\] 
and $\tilde C$ is of the same form as $C$ in \eqref{e:spec} but with $\lambda_i$ replaced by $\tilde\lambda_i=\lambda_i/(1-\lambda_i)$. See \citet{LMR12extended} and the references therein.

In general when dealing with DPPs on $\mathbb R^k$, the eigenfunctions in \eqref{e:spec} are unknown and \eqref{e:spec} needs to be replaced 
by an approximation. When $S$ is rectangular and the (unrestricted)
kernel is stationary, i.e., $C(\bx,\by)=C_0(\bx-\by)$,
\citet{LMR12extended,LMR15} discussed an approximation based on both the Fourier basis (used as eigenfunctions), the Fourier transform of $C_0$ on $\R^k$, a Fourier series expansion of $C_0$ on $S$, and a periodic extension of $C_0$ outside $S$.
This leads to infinite series approximations $\tilde C_{\mathrm{app}}$ and $D_{\mathrm{app}}$ of $\tilde C$ and $D$, respectively. In practice, a truncation of the infinite series is needed, leading to $\tilde C_{\mathrm{app},N}$ and $D_{\mathrm{app},N}$, say, where $N$ relates to the number of terms in the truncation and is increased until the approximate MLE stabilizes.  
 On the other hand, if $S=\mathbb S^d$ and $\bX$ is isotropic, 
the spectral representation \eqref{e:spec} is explicitly given in terms of known spherical harmonics, and so we do not an approximation except that a truncation is still used \citep[see][for details]{moeller:rubak:16}.  

A simulation study reported in \citet{LMR12extended} shows that approximate likelihood inference based on replacing $\tilde C$ by $\tilde C_{\mathrm{app},N}$ works well in practice.
As discussed in \citet{LMR12extended}, in case of stationary isotropic DPPs with a kernel of the form $C(\bx,\by)=\rho R_{0}(\|\bx-\by\|;\theta)$ where $(\rho,\theta)$ is the unknown parameter, the maximum likelihood estimate of the intensity $\rho$ is well approximated by the usual non-parametric estimate given by $\hat\rho=n/|S|$.
Further, \citet{LMR12extended,LMR15} discuss approximate MLE, model
comparison, and likelihood ratio tests for a number of examples of
specific data sets. They conclude that estimates obtained by moment
based methods as in Section~\ref{sec:est_intensities} give similar
results as those based on MLE, though they are somewhat less efficient. 
For non-stationary parametric DPP models, where the intensity depends on covariates while the pair correlation function is stationary, \citet{LMR12extended,LMR15} use the easier approach of minimum contrast estimation. 

%

\subsection{Bayesian inference} \label{sec:bayesinference}

Bayesian inference for parametric Poisson, Mat{\'e}rn
III hard core \citep{huber:wolpert:09,moeller:huber:wolpert:10},  Cox,
and cluster processes  have been reviewed in
\citet{moeller:waagepetersen:07} and \citet{guttorp:thorarinsdottir:12}. 
For Poisson and Mat{\'e}rn III models the likelihood is known. This is
also the case for Cox and cluster processes provided the latent random
function or the cluster centres are included among the unknown
parameters. Various hybrid (or Metropolis within Gibbs) algorithms for
posterior simulation apply, where e.g.\ in the case of cluster
processes the main ingredient is often a kind of birth-death-move Metropolis Hastings algorithm \citep{geyer:moeller:94,moeller:waagepetersen:03,huber:11}.
However, their implementations and a careful output analysis can be cumbersome, and relatively large computation times may be required. For a LGCP with a Mat{\'e}rn covariance function for the underlying Gaussian process,  
an alternative to MCMC is based on integrated nested Laplace approximations (INLA) \jmret{(see `Bayesian Computing with INLA: A Review' in this volume)}
provided certain restrictions are imposed on the smoothness parameter \citep{rue:martino:chopin:09,lindgren:rue:lindstroem:11,illian:soerbye:rue:12}. \citet{taylor:diggle:12}, however, question  
that INLA is always both significantly faster and more accurate than MCMC.

For a Gibbs point process, the unknown normalizing constant in the likelihood causes computational problems when calculating the Hastings ratio for MCMC posterior simulations; \citet{murray2006} call this `MCMC for doubly-intractable distributions'. \citet{moeller:pettitt:berthelsen:reeves:06} offer a solution based on an MCMC auxiliary variable method which involves perfect simulation \citep[many references to perfect simulation for spatial point processes can be found in][]{huber:15}.
In its simplest form, the auxiliary variable method may have low acceptance probabilities, but more elaborate methods were used in \citet{moeller:berthelsen:08} for pairwise interaction point processes, assuming that the first-order term is a shot noise process, and that the interaction function for a pair of points depends only on the distance between the two points and is a piecewise linear function. 
\citet{murray2006}'s exchange algorithm is a modification of the auxiliary variable method which is simpler to use.

\citet{guttorp:thorarinsdottir:12} discuss how to use Bayes factors for model selection problems, e.g.\ when considering a cluster process with different models for the dispersion distribution. The reversible jump MCMC algorithm \citep{green:95} is then used when proposing a jump from one dispersion distribution to another.     

\section{Functional summary statistics}\label{sec:summaries}

Classical functional summary statistics such as Ripley's $K$-function, the empty space function $F$, the nearest-neighbour function $G$, and the related $J$-function \citep[see e.g.][]{moeller:waagepetersen:03} play a major role in 
exploratory analysis of spatial point patterns and validation of
fitted models. For a
stationary point process, if $b(o,t)$ denotes the ball with center $o$ and radius $t$, $\rho K(t)= \mathrm E \#\{\bX^!_o\cap b(o,t)\}$ is the
expected number of further points in $\bX$ within distance $t$ of a 
typical point in $\bX$, while $F(t)=\mathrm P(\bX\cap b(o,t)\not=\emptyset)$ and
$G(t)=\mathrm P(\bX^!_o\cap b(o,t)\not=\emptyset)$ 
are probabilities of observing at least one
point within distance $t$ of respectively an arbitrary fixed point in space or a typical point in $\bX$. The $J$-function is defined as $J(t)=[1-G(t)]/[1-F(t)]$ for $t$
with $F(t)<1$. Section~\ref{s:newf} discusses some new summaries and
Section~\ref{s:enve} the use of envelopes and Monte Carlo tests.

\subsection{New functional summaries}\label{s:newf}

\cite{baddeley:moeller:waagepetersen:00} introduce the notion of
second order intensity-reweighted stationary point
processes and extend the definition of the $K$-function to such spatial
point processes. They briefly discuss ways of
generalizing $F$ and $G$ to non-stationary point processes but the practical
applicability of these generalizations was restricted to Poisson
processes. \cite{lieshout:11} is more successful in generalizing $F$,
$G$, and $J$. She defines the concept of intensity-reweighted moment
stationary (IRMS) point processes meaning that so-called $n$-point
correlation functions are translation invariant (whereby in particular
an intensity-reweighted moment stationary point process is also second
order intensity-reweighted stationary). In the stationary case, $F$,
$G$, and $J$ can be expressed in terms of series representations involving
joint intensities of all orders (provided the series are convergent)
and \cite{lieshout:11} extends these series representations to the
case of IRMS point processes. \cite{lieshout:11} further discusses specific examples of IMRS point
processes and non-parametric estimation of the generalized summary statistics.

To detect anisotropy in cases with linear structures such as in Figure~\ref{fig:columns_sphere} (left panel), \citet{moeller:safavimanesh:rasmusssen:15} introduce in the stationary and the second order intensity-reweighted stationary cases a functional summary statistic $K_{e}(r,t)$. This is called the cylindrical $K$-function since it is a directional
$K$-function whose structuring element
is a cylinder $C_e(r,t)$ centered at $o$ with direction $e$ (a unit vector), radius $r$, and height $2t$. In the stationary case, $\rho K_{e}(r,t)=\mathrm E\#\{\bX^!_o\cap C_e(r,t)\}$ is the
expected number of further points within the cylinder given that $\bX$ has a point at $o$. Choosing different directions and sizes of the cylinder, \citet{AliEtAl:16} and \citet{moeller:safavimanesh:rasmusssen:15} demonstrate that a non-parametric estimate of $K_{e}(r,t)$ is useful 
for detecting preferred directions and columnar structures in 2D and 3D  spatial point pattern data sets. 

For isotropic point processes on $\mathbb S^d$, \citet{robeson:li:huang:14} studied Ripley’s
$K$-function while \citet{lawrence:etal:16} and \citet{moeller:rubak:16} independently 
introduced empty space and nearest neighbour-functions $F$ and $G$ (and hence also a $J$-function). The interpretations are similar as for stationary point processes on $\mathbb R^k$ but using great circle distance on the sphere (the focus in the papers are on the case $d=2$ but most theory easily extends to the general case of dimension $d=1,2,\ldots$). In the case of second order intensity-reweighted isotropy, 
\citet{lawrence:etal:16} and \citet{moeller:rubak:16}
 also study the inhomogeneous $K$-function. While \citet{moeller:rubak:16} provide the technical details on how to define Palm distributions for point processes on $\mathbb S^d$ and illustrate the application of functional summary statistics for DPPs on the sphere, \citet{lawrence:etal:16} deal with edge effects and a cluster point process on the sphere used for modelling the data set in Figure~\ref{fig:columns_sphere} (right panel).  

\subsection{Envelopes and tests}\label{s:enve}

A functional summary statistic such as $K(r)$ contains information from different spatial scales. Usually we consider a graphical representation of a non-parametric estimator of e.g.\ $K$ together with a simulation envelope which expresses the variability of this estimator. Below we discuss how formal statistical tests may be constructed from such an envelope.

Suppose we have observed a realization of a spatial point process $\bX_1$ and then simulated spatial point process realizations $\bX_2,\ldots,\bX_{m+1}$ under a claimed model for $\bX_1$ so that the joint distribution of $\bX_1,\ldots,\bX_{m+1}$ is exchangeable under the model. For $i=1,\ldots,m+1$, let $\widehat T_i(r)=\widehat T(r,\bX_i)$ ($r>0$) denote an estimator of a functional summary statistic such as $F,G,J,K$ or their inhomogeneous versions. 
For each value of $r>0$ and $k=1,2,\ldots$, we define a {\it pointwise envelope} with lower and upper bounds $\widehat T^{(k)}_{\mathrm{low}}(r)$ and $\widehat T^{(k)}_{\mathrm{up}}(r)$ given by the $k$th smallest and largest values of $\widehat T_1(r),\ldots,\widehat T_{m+1}(r)$. Then $\widehat T_1(r)$ is within $\widehat T^{(k)}_{\mathrm{low}}(r)$ and $\widehat T^{(k)}_{\mathrm{up}}(r)$ with probability $\alpha:=2k/(m+1)$ or in case of ties with approximate probability $\alpha$. Thus a plot of the pointwise envelopes over a range of $r$ values enables us to assess {\it for each fixed value} of $r$ a Monte Carlo test where at (approximate) level $\alpha$ we reject the claimed model if $\widehat T_1(r)$ is outside the envelope. If relevant, this may be replaced by a one-sided Monte Carlo test \citep[see e.g.][Section 10.7.5]{baddeley:rubak:turner:15}.  

Several authors have warned against the interpretation of the pointwise envelope and the corresponding Monte Carlo test. In practice, the specified model for the data is an estimated model under a composite hypothesis, and so the Monte Carlo test is strictly speaking invalid but usually conservative. A global envelope test can be obtained by rejecting the claimed model if $\widehat T_1(r)$ is not always inside the pointwise envelopes for a given finite selection of $r$-values. However, \citet{ripley:77} noticed that due to multiple testing this gives an unknown probability of committing a type I error which is larger than the level $\af$ associated with each of the pointwise tests. See also \citet{loosmore:ford:06} and 
\citet[Section 10.7.2]{baddeley:rubak:turner:15}. 

To solve the multiple testing problem, ways of constructing $p$-values corresponding to a {\it global envelope} with lower and upper bounds $(\widehat T_{\mathrm{low}}(r))_{r\in I}$ and $(\widehat T_{\mathrm{up}}(r))_{r\in I}$ for 
a given finite set $I\subset(0,\infty)$ (approximating a predefined interval of $r$-values) and number $m$ of simulations have been suggested in \citet{myllymaki:etal:15}; see also \citet[Chapter 10]{baddeley:rubak:turner:15}. In one approach, a so-called rank envelope test is based on extreme ranks $R_1,\ldots,R_{m+1}$, where $R_i$ is the largest $k$ so that  $\widehat T_i(r)$ is within $\widehat T^{(k)}_{\mathrm{low}}(r)$ and $\widehat T^{(k)}_{\mathrm{up}}(r)$ for all $r\in I$, cf.\  
\citet{myllymaki:etal:15}.
They notice that the extreme ranks are tied and 
 show how a liberal/lower $p$-value $p_{-}$ and a conservative/upper $p$-value $p_{+}$ can be constructed.  For a given probability $\af$, liberal/conservative refers to the test obtained by rejecting if respectively $p_{-}$ or $p_{+}$ is less or equal to $\alpha$. \citet{myllymaki:etal:15} further define the $100(1-\alpha)$\% rank envelope by the bounds $\widehat T^{(k_\alpha)}_{\mathrm{low}}(r)$ and $\widehat T^{(k_\alpha)}_{\mathrm{up}}(r)$ for $r\in I$, where $k_\alpha$ is
 the largest $k$ such that $\#\{R_i<k\}\le\alpha(m+1)$. They show that rejecting due to $p_{+} \le \alpha$ is equivalent to that $\widehat T_1(r)$ is strictly outside the global rank envelope for at least one $r \in I$.  They also show how to use a more
comprehensive summary of the pointwise ranks, namely so-called rank counts. 
 Regarding the number of simulations, \citet{myllymaki:etal:15} recommend to use at least $m=2500$ when $\af=5$\%.

Finally, the approach in \citet{dao:genton:14} to reduce or eliminate the problem of conservatism  has been adapted in \citet{myllymaki:etal:15} and \citet[Section 10.10]{baddeley:rubak:turner:15}. Software for the methods in \citet{myllymaki:etal:15} is provided as an R library spptest  at
https://github.com/myllym/spptest and it will be available in  
\texttt{spatstat}.
  
\section{Further recent developments and perspectives for future research}\label{sec:future}

We conclude with a brief discussion on some other recent and future developments and some open problems. 

In practice `spatial' often refers to two dimensions, however
3D point pattern data sets are becoming more
common, e.g.\ as illustrated in connection to the pyramidal brain cell data set (Figure~\ref{fig:columns_sphere})  and in \cite{khanmohammadi:waagepetersen:sporring:15} who
consider 3D marked point pattern data obtained from focused ion beam
scanning electron microscopial  images. Most methods for analyzing
spatial point pattern data sets are from a theoretical point of view applicable
in any spatial dimension. From a practical point of view, however, the
3D case poses additional computational challenges. This
may again require developments of the existing theory to ensure practically applicable methods of analysis.

Due to space constraints we have not covered developments for marked
and multivariate spatial point processes. Research in multivariate
point processes has to a large extent focused on the bivariate or trivariate \cite[e.g.\ ][]{baddeley:jammalamadaka:nair:14} case
but in ecology large data sets with locations of thousands of trees for hundreds of species are collected. Such data sets call for research in methods for
analysing highly multivariate
point patterns. Steps in this direction are taken in
\cite{jalilian:guan:mateu:waagepetersen:15} and \cite{waagepetersen:guan:jalilian:mateu:16}.  
However, considerable challenges remain in order to obtain practically applicable statistical models and methods for point patterns with hundreds of types of points.

We have also not covered point processes defined on a network of lines, e.g.\ in connection to road accidents or spines on the dendrite networks of a neuron \citep{baddeley:jammalamadaka:nair:14,baddeley:rubak:turner:15}. Here statistical models and methods should take into account the network geometry as addressed in \cite{ang:baddeley:nair:12}, \cite{okabe:sugihara:12}, and the references therein. Assuming that 
the pair correlation function only depends on shortest path distance, \cite{ang:baddeley:nair:12} show how to define the counterpart of Ripley's $K$-function. There is a lack of models with this property apart from the Poisson process and another specific model studied in \cite{ang:baddeley:nair:12}. Currently, together with Ethan Anderes and Jakob G.\ Rasmussen, the first author of the present paper is developing covariance functions on linear networks which only depend on shortest path distance, and thereby DPPs and LGCPs may be constructed on such networks so that 
the pair correlation function only depends on shortest path distance.  


As discussed in Section~\ref{s:latent}, 
many observed spatial point patterns contain points placed
roughly on line segments. However, in some cases
the exact mechanism responsible for the formations of lines is
unknown. For instance, \cite{moeller:rasmussen:12} model linear structures for locations  of  bronze
age graves in Denmark  and mountain  tops  in  Spain, without introducing a latent line segment process. Instead they use sequential point process models,
i.e.\ where the points are ordered \citep{lieshout:06a,lieshout:06b}. Since the ordering is not known it is treated as a latent variable. 

Estimating functions for spatial point processes can also be derived using a variational approach inspired by the variational estimators for Markov random fields developed by \cite{almeida:gidas:93}.  \cite{baddeley:dereudre:13} do this for Gibbs point processes, while \cite{coeurjolly:moeller:14} consider another variational estimator for the intensity function. In comparison with the estimation procedures in Section~\ref{s:compositelikelihood}, the variational estimators are exact and explicit, quicker to use, and very simple to implement. However, inference is
effectively conditional on the observed number of points, and so the intensity parameter cannot be estimated. In particular the variational estimator in \cite{coeurjolly:moeller:14} is simple to use, but there may be a loss
in efficiency, since it does not take into account spatial correlation or interaction. 
\cite{baddeley:dereudre:13} and \cite{coeurjolly:moeller:14}  establish 
strong consistency and asymptotic normality of their variational estimators, and they also discuss finite sample
properties in comparison to the estimation methods in Section~\ref{s:compositelikelihood}.

 \cite{coeurjolly:guan:khanmohammadi:waagepetersen:15} follow ideas in \cite{guan:jalilian:waagepetersen:15} to construct a
weight function that in certain situations outperforms the weight function for the
pseudo-likelihood. However, it is still not known what is the optimal weight function in \eqref{eq:etf}. 
To the best of our knowledge, MLE for inhomogeneous DPPs has yet not be studied. Seemingly, it also remains to consider MLE for the case $S=\mathbb S^d$.

While \texttt{spatstat} supports a wealth of inference procedures based on frequentist methods, including most of those discussed in Sections~\ref{sec:estimation}-\ref{sec:summaries}, \texttt{spatstat} is so far of very limited use for Bayesian inference. 
Instead various software packages for Bayesian analysis have been developed in connection to specific Poisson, cluster, Cox and Gibbs point process models. Moreover, as far as we know Bayesian analysis for DPPs is another unexplored area. 


\section*{DISCLOSURE STATEMENT}
The authors are not aware of any affiliations, memberships, funding, or financial holdings that
might be perceived as affecting the objectivity of this review. 

\section*{ACKNOWLEDGMENTS}
Supported by the Danish Council for Independent Research | Natural
Sciences, grant 12-124675, "Mathematical and Statistical Analysis of Spatial Data", and by
the "Centre for Stochastic Geometry and Advanced Bioimaging", funded by grant 8721 from the Villum Foundation.

%
\bibliographystyle{ar-style1}
\bibliography{masterbib}

\begin{thebibliography}{}
\expandafter\ifx\csname natexlab\endcsname\relax\def\natexlab#1{#1}\fi

\bibitem[Almeida \& Gidas(1993)]{almeida:gidas:93}
Almeida MP, Gidas B. 1993.
A variational method for estimating the parameters of {MRF} from complete or
  incomplete data.
\textit{Annals of Applied Probability} 3:103--136

\bibitem[Andersen \& Hahn(2015)]{andersen:hahn:15}
Andersen IT, Hahn U. 2015.
{Mat{\'e}rn thinned {C}ox processes}.
\textit{Spatial Statistics} To appear

\bibitem[Ang et~al.(2012)Ang, Baddeley \& Nair]{ang:baddeley:nair:12}
Ang QW, Baddeley A, Nair G. 2012.
Geometrically corrected second order analysis of events on a linear network,
  with applications to ecology and criminology.
\textit{Scandinavian Journal of Statistics} 39:591–--617

\bibitem[Baddeley \& Dereudre(2013)]{baddeley:dereudre:13}
Baddeley A, Dereudre D. 2013.
Variational estimators for the parameters of {G}ibbs point process models.
\textit{Bernoulli} 19:905--930

\bibitem[Baddeley et~al.(2014{\natexlab{a}})Baddeley, Jammalamadaka \&
  Nair]{baddeley:jammalamadaka:nair:14}
Baddeley A, Jammalamadaka A, Nair G. 2014{\natexlab{a}}.
Multitype point process analysis of spines on the dendrite network of a neuron.
\textit{Journal of the Royal Statistical Society: Series C (Applied
  Statistics)} 63:673–--694

\bibitem[Baddeley \& Nair(2012{\natexlab{a}})]{baddeley:nair:12b}
Baddeley A, Nair G. 2012{\natexlab{a}}.
Approximating the moments of a spatial point process.
\textit{Stat} 1:18--30

\bibitem[Baddeley \& Nair(2012{\natexlab{b}})]{baddeley:nair:12a}
Baddeley A, Nair G. 2012{\natexlab{b}}.
Fast approximation of the intensity of {G}ibbs point processes.
\textit{Electronic Journal of Statistics} 6:1155--1169

\bibitem[Baddeley \& Turner(2000)]{baddeley:turner:00}
Baddeley A, Turner R. 2000.
Practical maximum pseudolikelihood for spatial point patterns.
\textit{Australian and New Zealand Journal of Statistics} 42:283--322

\bibitem[Baddeley \& Turner(2005)]{baddeley:turner:05}
Baddeley A, Turner R. 2005.
Spatstat: an {{\sf R}} package for analyzing spatial point patterns.
\textit{Journal of Statistical Software} 12:1--42.
{URL}: {{\tt www.jstatsoft.org}}, {ISSN}: 1548-7660

\bibitem[Baddeley \& Turner(2006)]{baddeley:turner:06}
Baddeley A, Turner R. 2006.
Modelling spatial point patterns in r. In \textit{Case Studies in Spatial Point
  Pattern Modelling, number 185 in Lecture Notes in Statistics}, eds.
  A~Baddeley, P~Gregori, J~Mateu, R~Stoica, D~Stoyan. New York:
  Springer-Verlag,  23--74

\bibitem[Baddeley et~al.(2014{\natexlab{b}})Baddeley, Couerjolly, Rubak \&
  Waagepetersen]{baddeley:etal:14}
Baddeley AJ, Couerjolly JF, Rubak E, Waagepetersen R. 2014{\natexlab{b}}.
Logistic regression for spatial {G}ibbs point processes.
\textit{Biometrika} 101:377--392

\bibitem[Baddeley et~al.(2000)Baddeley, M{\o}ller \&
  Waagepetersen]{baddeley:moeller:waagepetersen:00}
Baddeley AJ, M{\o}ller J, Waagepetersen R. 2000.
Non- and semi-parametric estimation of interaction in inhomogeneous point
  patterns.
\textit{Statistica Neerlandica} 54:329--350

\bibitem[Baddeley et~al.(2015)Baddeley, Rubak \&
  Turner]{baddeley:rubak:turner:15}
Baddeley AJ, Rubak E, Turner R. 2015.
Spatial point patterns: Methodology and applications with {R}.
Interdisciplinary Statistics. Boca Raton, Florida: Chapman \& Hall/CRC

\bibitem[Baddeley et~al.(2005)Baddeley, Turner, M{\o}ller \&
  Hazelton]{baddeley:etal:05}
Baddeley AJ, Turner R, M{\o}ller J, Hazelton M. 2005.
Residual analysis for spatial point processes (with discussion).
\textit{Journal of the Royal Statistical Society, Series B} 67:617--666

\bibitem[Berman(1986)]{berman:86}
Berman M. 1986.
Testing for spatial association between a point process and another stochastic
  process.
\textit{Applied Statistics} 35:54--62

\bibitem[Berman \& Turner(1992)]{berman:turner:92}
Berman M, Turner R. 1992.
Approximating point process likelihoods with {G}{L}{I}{M}.
\textit{Applied Statistics} 41:31--38

\bibitem[Berthelsen \& M{\o}ller(2006)]{moeller:berthelsen:08}
Berthelsen K, M{\o}ller J. 2006.
Non-parametric {B}ayesian inference for inhomogeneous {M}arkov point processes.
\textit{Australian and New Zealand Journal of Statistics} 50:257--272

\bibitem[Besag(1977)]{besag:77}
Besag J. 1977.
Some methods of statistical analysis for spatial data.
\textit{Bulletin of the International Statistical Institute} 47:77--92

\bibitem[Billiot(1997)]{billiot:97}
Billiot JM. 1997.
Asymptotic properties of {T}akacs-{F}iksel estimation method for {G}ibbs point
  processes.
\textit{Statistics} 30:68--89

\bibitem[Billiot et~al.(2008)Billiot, Coeurjolly \&
  Drouilhet]{billiot:coeurjolly:drouilhet:08}
Billiot JM, Coeurjolly JF, Drouilhet R. 2008.
{M}aximum pseudolikelihood estimator for exponential family models of marked
  {G}ibbs point processes.
\textit{Electronic Journal of Statistics} 2:234--264

\bibitem[Blackwell(2001)]{blackwell:01}
Blackwell P. 2001.
Bayesian inference for a random tessellation process.
\textit{Biometrics} 57:502--507

\bibitem[Blackwell \& M{\o}ller(2003)]{blackwell:moeller:02}
Blackwell P, M{\o}ller J. 2003.
Bayesian analysis of deformed tessellation models.
\textit{Advances in Applied Probability} 35:4--26

\bibitem[Chang et~al.(2015)Chang, Waagepetersen, Yu, Ma, Holford
  et~al.]{chang:etal:15}
Chang X, Waagepetersen R, Yu H, Ma X, Holford T, et~al. 2015.
Disease risk estimation by combining case-control data with aggregated
  information on the population at risk.
\textit{Biometrics} 71:114--121

\bibitem[Chiu et~al.(2013)Chiu, Stoyan, Kendall \&
  Mecke]{chiu:stoyan:kendall:mecke:13}
Chiu SN, Stoyan D, Kendall WS, Mecke J. 2013.
Stochastic geometry and its applications.
Wiley, Chichester, 3rd ed.

\bibitem[Coeurjolly et~al.(2012)Coeurjolly, Dereudre, Drouilhet \&
  Lavancier]{coeurjolly:dereudre:lavancier:12}
Coeurjolly JF, Dereudre D, Drouilhet R, Lavancier F. 2012.
{T}akacs-{F}iksel method for stationary marked {G}ibbs point processes.
\textit{Scandinavian Journal of Statistics} 49:416--443

\bibitem[Coeurjolly et~al.(2015{\natexlab{a}})Coeurjolly, Guan, Khanmohammadi
  \& Waagepetersen]{coeurjolly:guan:khanmohammadi:waagepetersen:15}
Coeurjolly JF, Guan Y, Khanmohammadi M, Waagepetersen R. 2015{\natexlab{a}}.
Towards optimal {T}akacs-{F}iksel estimation.
CSGB Research Reports~15, Centre for Stochastic Geometry and Advanced
  Bioimaging.
Submitted for journal publication

\bibitem[Coeurjolly \& M{\o}ller(2014)]{coeurjolly:moeller:14}
Coeurjolly JF, M{\o}ller J. 2014.
Variational approach for spatial point process intensity estimation.
\textit{Bernoulli} 20:1097--1125

\bibitem[Coeurjolly et~al.(2015{\natexlab{b}})Coeurjolly, M{\o}ller \&
  Waagepetersen]{coeurjolly:moeller:waagepetersen:tutorial}
Coeurjolly JF, M{\o}ller J, Waagepetersen R. 2015{\natexlab{b}}.
Conditioning in spatial point processes.
CSGB Research Reports~14, Centre for Stochastic Geometry and Advanced
  Bioimaging.
Submitted for publication. Preprint on arXiv: 1512.05871

\bibitem[Condit(1998)]{condit:98}
Condit R. 1998.
Tropical forest census plots.
Berlin, Germany and Georgetown, Texas: Springer-Verlag and R.\ G.\ Landes
  Company

\bibitem[Condit et~al.(1996)Condit, Hubbell \&
  Foster]{condit:hubbell:foster:96}
Condit R, Hubbell SP, Foster RB. 1996.
Changes in tree species abundance in a neotropical forest: impact of climate
  change.
\textit{Journal of Tropical Ecology} 12:231--256

\bibitem[Dao \& Genton(2014)]{dao:genton:14}
Dao NA, Genton MG. 2014.
A {M}onte {C}arlo adjusted goodness-of-fit test for parametric models
  describing spatial point patterns.
\textit{Journal of Computational and Graphical Statistics} 23:497--517

\bibitem[Deng et~al.(2015)Deng, Waagepetersen \&
  Guan]{deng:waagepetersen:guan:14}
Deng C, Waagepetersen R, Guan Y. 2015.
A combined estimating function approach for fitting stationary point process
  models.
\textit{Biometrika} 101:393--408

\bibitem[Diggle et~al.(1994)Diggle, Fiksel, Grabarnik, Ogata, Stoyan \&
  Tanemura]{diggle:etal:94}
Diggle P, Fiksel T, Grabarnik P, Ogata Y, Stoyan D, Tanemura M. 1994.
On parameter estimation for pairwise interaction point processes.
\textit{International Statistical Review} 62:99--117

\bibitem[Diggle et~al.(2010)Diggle, Guan, Hart, Paize \&
  Stanton]{diggle:etal:10}
Diggle P, Guan Y, Hart A, Paize F, Stanton M. 2010.
Estimating individual-level risk in spatial epidemiology using spatially
  aggregated information on the population at risk.
\textit{Journal of the American Statistical Association} 105:1394--1402

\bibitem[Diggle(2003)]{diggle:03}
Diggle PJ. 2003.
Statistical analysis of spatial point patterns.
London: Arnold, 2nd ed.

\bibitem[Diggle \& Rowlingson(1994)]{diggle:rowlingson:94}
Diggle PJ, Rowlingson B. 1994.
A conditional approach to point process modelling of elevated risk.
\textit{Journal of the Royal Statistical Society, Series A (Statistics in
  Society)} 157:433--440

\bibitem[Dvo{\v r}{\' a}k \& Proke{\v s}ov{\' a}(2012)]{dvorak:prokesova:12}
Dvo{\v r}{\' a}k J, Proke{\v s}ov{\' a} M. 2012.
Moment estimation methods for stationary spatial {C}ox processes - a
  comparison.
\textit{Kybernetika} :1007--1026

\bibitem[Fiksel(1984)]{fiksel:84}
Fiksel T. 1984.
Estimation of parameterized pair potentials of marked and nonmarked {G}ibbsian
  point processes.
\textit{Elektronische Informationsverarbeitung und Kypernetik} 20:270--278

\bibitem[Gelfand et~al.(2010)Gelfand, Diggle, Guttorp \&
  Fuentes]{gelfand2010handbook}
Gelfand A, Diggle P, Guttorp P, Fuentes M. 2010.
Handbook of spatial statistics.
Chapman \& Hall/CRC Handbooks of Modern Statistical Methods. Taylor \& Francis

\bibitem[Georgii(1976)]{georgii:76}
Georgii HO. 1976.
{{C}anonical and grand canonical {G}ibbs states for continuum systems}.
\textit{Communications in Mathematical Physics} 48:31--51

\bibitem[Geyer \& M{\o}ller(1994)]{geyer:moeller:94}
Geyer CJ, M{\o}ller J. 1994.
Simulation procedures and likelihood inference for spatial point processes.
\textit{Scandinavian Journal of Statistics} 21:359--373

\bibitem[Goldstein et~al.(2015)Goldstein, Haran, Simeonov, Fricks \&
  Chiaromonte]{goldstein2014}
Goldstein J, Haran M, Simeonov I, Fricks J, Chiaromonte F. 2015.
An attraction-repulsion point process model for respiratory syncytial virus
  infections.
\textit{Biometrics} To appear

\bibitem[Green(1995)]{green:95}
Green P. 1995.
Reversible jump {M}arkov chain {M}onte {C}arlo computation and {B}ayesian model
  determination.
\textit{Biometrika} 82:711--732

\bibitem[Guan(2006)]{guan:06}
Guan Y. 2006.
A composite likelihood approach in fitting spatial point process models.
\textit{Journal of the American Statistical Association} 101:1502--1512

\bibitem[Guan \& Loh(2007)]{guan:loh:07}
Guan Y, Loh JM. 2007.
A thinned block bootstrap procedure for modeling inhomogeneous spatial point
  patterns.
\textit{Journal of the American Statistical Association} 102:1377--1386

\bibitem[Guan et~al.(2008)Guan, Waagepetersen \&
  Beale]{guan:waagepetersen:beale:08}
Guan Y, Waagepetersen R, Beale CM. 2008.
Second-order analysis of inhomogeneous spatial point processes with
  proportional intensity functions.
\textit{Journal of the American Statistical Association} 103:769--777

\bibitem[Guan et~al.(2015)Guan, Waagepetersen \&
  Jalilian]{guan:jalilian:waagepetersen:15}
Guan Y, Waagepetersen R, Jalilian A. 2015.
Quasi-likelihood for spatial point processes.
\textit{Journal of the Royal Statistical Society, Series B} 77:677--697

\bibitem[Guan et~al.(2016)Guan, Zhang \& Waagepetersen]{guan:etal:16}
Guan Y, Zhang E, Waagepetersen R. 2016.
Second-order quasi-likelihood for spatial point processes.
Submitted

\bibitem[Guttorp \& Thorarinsdottir(2012)]{guttorp:thorarinsdottir:12}
Guttorp P, Thorarinsdottir T. 2012.
Bayesian inference for non-{M}arkovian point processes. In \textit{Advances and
  Challenges in Space-time Modelling of Natural Events}, eds. E~Porcu,
  J~Montero, M~Schlather, Lecture Notes in Statistics. Springer: Berlin
  Heidelberg,  79--102

\bibitem[Hahn et~al.(2003)Hahn, Jensen, van Lieshout \& Nielsen]{hahn2003}
Hahn U, Jensen EBV, van Lieshout MNM, Nielsen LS. 2003.
Inhomogeneous spatial point processes by location-dependent scaling.
\textit{Advances in Applied Probability} 35:319--336

\bibitem[Heyde(1997)]{heyde:97}
Heyde CC. 1997.
Quasi-likelihood and its application - a general approach to optimal parameter
  estimation.
Springer Series in Statistics. Springer

\bibitem[Hodder \& Orton(2013)]{hodder:orton:76}
Hodder I, Orton C. 2013.
Spatial {A}nalysis in {A}rchaeology.
Cambridge University Press, Cambridge

\bibitem[Hough et~al.(2006)Hough, Krishnapur, Peres \&
  Vir\`{a}g]{Hough:etal:06}
Hough JB, Krishnapur M, Peres Y, Vir\`{a}g B. 2006.
Determinantal processes and independence.
\textit{Probability Surveys} 3:206--229

\bibitem[Huang et~al.(2014)Huang, Ma, Waagepetersen, Holford, Wang
  et~al.]{huang:etal:14}
Huang H, Ma X, Waagepetersen R, Holford T, Wang R, et~al. 2014.
A new estimation approach for combining epidemiological data from multiple
  sources.
\textit{Journal of the American Statistical Association} 109:11--23

\bibitem[Hubbell \& Foster(1983)]{hubbell:foster:83}
Hubbell SP, Foster RB. 1983.
Diversity of canopy trees in a neotropical forest and implications for
  conservation. In \textit{Tropical Rain Forest: Ecology and Management}, eds.
  SL~Sutton, TC~Whitmore, AC~Chadwick. Oxford: Blackwell Scientific
  Publications,  25--41

\bibitem[Huber(2011)]{huber:11}
Huber M. 2011.
Spatial point processes. In \textit{Handbook of MCMC}, eds. S~Brooks, A~Gelman,
  G~Jones, X~Meng. Chapman \& Hall/CRC Press,  227--252

\bibitem[Huber(2015)]{huber:15}
Huber M. 2015.
Perfect simulation.
Chapman \& Hall/CRC, Boca Raton

\bibitem[Huber \& Wolpert(2009)]{huber:wolpert:09}
Huber M, Wolpert R. 2009.
Likelihood-based inference for {M}at\'{e}rn type-{III} repulsive point
  processes.
\textit{Advances in Applied Probability} 41:958--977

\bibitem[Icke \& {V}an~de Weygaert(1987)]{icke:weygaert:87}
Icke V, {V}an~de Weygaert R. 1987.
Fragmenting the {U}niverse.
\textit{Astronomy and Astrophysics} 184:16--32

\bibitem[Illian et~al.(2008)Illian, Penttinen, Stoyan \&
  Stoyan]{illian:etal:08}
Illian J, Penttinen A, Stoyan H, Stoyan D. 2008.
Statistical analysis and modelling of spatial point patterns.
Statistics in Practice. New York: Wiley

\bibitem[Illian et~al.(2012)Illian, S{\o}rbye \& Rue]{illian:soerbye:rue:12}
Illian JB, S{\o}rbye SH, Rue H. 2012.
A toolbox for fitting complex spatial point process models using integrated
  nested {L}aplace approximation ({INLA}).
\textit{The Annals of Applied Statistics} 6:1499--1530

\bibitem[Jalilian et~al.(2015)Jalilian, Guan, Mateu \&
  Waagepetersen]{jalilian:guan:mateu:waagepetersen:15}
Jalilian A, Guan Y, Mateu J, Waagepetersen R. 2015.
Multivariate product-shot-noise {C}ox point process models.
\textit{Biometrics} 71:1022--1033

\bibitem[Jalilian et~al.(2013)Jalilian, Guan \&
  Waagepetersen]{jalilian:guan:waagepetersen:13}
Jalilian A, Guan Y, Waagepetersen R. 2013.
Decomposition of variance for spatial {C}ox processes.
\textit{Scandinavian Journal of Statistics} 40:119--137

\bibitem[Jensen \& Nielsen(2000)]{jensen:nielsen:00}
Jensen EBV, Nielsen LS. 2000.
Inhomogeneous {M}arkov point processes by transformation.
\textit{Bernoulli} 6:761--782

\bibitem[Jensen \& M{\o}ller(1991)]{jensen:moeller:91}
Jensen JL, M{\o}ller J. 1991.
Pseudolikelihood for exponential family models of spatial point processes.
\textit{Annals of Applied Probability} 3:445--461

\bibitem[Karlsson \& Liljeborg(1994)]{karlsson:liljeborg:94}
Karlsson L, Liljeborg A. 1994.
Second-order stereology for pores in translucent alumina studied by confocal
  scanning laser microscopy.
\textit{Journal of Microscopy} 175:186--194

\bibitem[Kendall et~al.(1999)Kendall, van Lieshout \&
  Baddeley]{kendall:lieshout:baddeley:99}
Kendall WS, van Lieshout MNM, Baddeley AJ. 1999.
Quermass-interaction processes: conditions for stability.
\textit{Advances in Applied Probability} 31:315--342

\bibitem[Khanmohammadi et~al.(2014)Khanmohammadi, Waagepetersen, Nava,
  Nyengaard \& Sporring]{khanmohammadi:etal:14}
Khanmohammadi M, Waagepetersen R, Nava N, Nyengaard JR, Sporring J. 2014.
Analysing the distribution of synaptic vesicles using a spatial point process
  model, In \textit{Proceedings of the 5th {ACM} Conference on Bioinformatics,
  Computational Biology, and Health Informatics, {BCB} '14, Newport Beach,
  California, USA, September 20-23, 2014}

\bibitem[Khanmohammadi et~al.(2015)Khanmohammadi, Waagepetersen \&
  Sporring]{khanmohammadi:waagepetersen:sporring:15}
Khanmohammadi M, Waagepetersen R, Sporring J. 2015.
Analysis of shape and spatial interaction of synaptic vesicles using data from
  focused ion beam scanning electron microscopy ({FIB-SEM}).
\textit{Frontiers in Neuroanatomy} 9

\bibitem[Lavancier \& M{\o}ller(2016)]{lavancier:moeller:16}
Lavancier F, M{\o}ller J. 2016.
Modelling aggregation on the large scale and regularity on the small scale in
  spatial point pattern datasets.
\textit{Scandinavian Journal of Statistics} 43:587--609

\bibitem[Lavancier et~al.(2014)Lavancier, M{\o}ller \& Rubak]{LMR12extended}
Lavancier F, M{\o}ller J, Rubak E. 2014.
Determinantal point process models and statistical inference (extended
  version).
Preprint on arxiv: 1205.4818

\bibitem[Lavancier et~al.(2015)Lavancier, M{\o}ller \& Rubak]{LMR15}
Lavancier F, M{\o}ller J, Rubak E. 2015.
Determinantal point process models and statistical inference.
\textit{Journal of the Royal Statistical Society: Series B (Statistical
  Methodology)} 77:853--877

\bibitem[Lavancier \& Rochet(2016)]{lavancier:rochet:16}
Lavancier F, Rochet P. 2016.
A general method to combine estimators.
\textit{Computational Statistics and Data Analysis} 94:175--192

\bibitem[Lawrence et~al.(2016)Lawrence, Baddeley, Milne \&
  Nair]{lawrence:etal:16}
Lawrence T, Baddeley A, Milne RK, Nair G. 2016.
Point pattern analysis on a region of a sphere.
\textit{Stat} 5:144--157

\bibitem[Lindgren et~al.(2011)Lindgren, Rue \& Lindstr{\"
  o}m]{lindgren:rue:lindstroem:11}
Lindgren F, Rue H, Lindstr{\" o}m J. 2011.
An explicit link between {G}aussian fields and {G}aussian {M}arkov random
  fields: the stochastic partial differential equation approach.
\textit{Journal of the Royal Statistical Society, Series B} 73:423--498

\bibitem[Loosmore \& Ford(2006)]{loosmore:ford:06}
Loosmore N, Ford E. 2006.
Statistical inference using the {$G$} or {$K$} point pattern spatial
  statistics.
\textit{Ecology} 87:1925–--1931

\bibitem[Macchi(1975)]{macchi:75}
Macchi O. 1975.
The coincidence approach to stochastic point processes.
\textit{Advances in Applied Probability} 7:83--122

\bibitem[Mat\'{e}rn(1986)]{matern:86}
Mat\'{e}rn B. 1986.
Spatial variation.
No.~36 in Lecture Notes in Statistics. Berlin: Springer-Verlag

\bibitem[McCullagh \& M{\o}ller(2006)]{mccullagh:moeller:05}
McCullagh P, M{\o}ller J. 2006.
The permanental process.
\textit{Advances in Applied Probability} 38:873--888

\bibitem[McKeague \& Loizeaux(2002)]{mckeague:loizeaux:02}
McKeague IW, Loizeaux M. 2002.
Perfect sampling for point process cluster modelling. In \textit{Spatial
  Cluster Modelling}, eds. AB~Lawson, D~Denison. Boca Raton, Florida: Chapman
  \& Hall/CRC Press,  87--107

\bibitem[M{\o}ller(1999)]{moeller:97aa}
M{\o}ller J. 1999.
Markov chain {M}onte {C}arlo and spatial point processes. In \textit{Stochastic
  Geometry: Likelihood and Computation}, eds. OE~Barndorff-Nielsen, WS~Kendall,
  MNM van Lieshout. Boca Raton: Monographs on Statistics and Applied
  Probability 80, Chapman and Hall/CRC,  141--172

\bibitem[M{{\o}}ller(2003)]{moeller:03}
M{{\o}}ller J. 2003.
Shot noise {C}ox processes.
\textit{Advances in Applied Probability} 35:614--640

\bibitem[M{\o}ller et~al.(2010)M{\o}ller, Huber \&
  Wolpert]{moeller:huber:wolpert:10}
M{\o}ller J, Huber M, Wolpert R. 2010.
Perfect simulation and moment properties for the {M}at\'{e}rn type {III}
  process.
\textit{Stochastic Processes and their Applications} 120:2142--2158

\bibitem[M{\o}ller et~al.(2015{\natexlab{a}})M{\o}ller, Nielsen, Porcu \&
  Rubak]{moeller:nielsen:porcu:rubak:15}
M{\o}ller J, Nielsen M, Porcu E, Rubak E. 2015{\natexlab{a}}.
Determinantal point process models on the sphere.
CSGB {R}esearch {R}eports~13, {Centre for Stochastic Geometry and Advanced
  Bioimaging}.
Preprint on arXiv: 1607.03675. To appear in Bernoulli

\bibitem[M{\o}ller et~al.(2006)M{\o}ller, Pettitt, Berthelsen \&
  Reeves]{moeller:pettitt:berthelsen:reeves:06}
M{\o}ller J, Pettitt A, Berthelsen K, Reeves R. 2006.
An efficient {M}arkov chain {M}onte {C}arlo method for distributions with
  intractable normalising constants.
\textit{Biometrika} 93:451--458

\bibitem[M{\o}ller \& Rasmussen(2015)]{moller:rasmussen:15}
M{\o}ller J, Rasmussen J. 2015.
Spatial cluster point processes related to {P}oisson-{V}oronoi tessellations.
\textit{Stochastic Environmental Research and Risk Assessment} 29:431--441

\bibitem[M{\o}ller \& Rasmussen(2012)]{moeller:rasmussen:12}
M{\o}ller J, Rasmussen JG. 2012.
A sequential point process model and {B}ayesian inference for spatial point
  patterns with linear structures.
\textit{Scandinavian Journal of Statistics} 39:618--634

\bibitem[M{\o}ller \& Rubak(2016)]{moeller:rubak:16}
M{\o}ller J, Rubak E. 2016.
Determinantal point processes and functional summary statistics on the sphere.
CSGB {R}esearch {R}eports~2, {Centre for Stochastic Geometry and Advanced
  Bioimaging}.
Preprint on arXiv: 1601.03448. To appear in Spatial Statistics

\bibitem[M{\o}ller et~al.(2015{\natexlab{b}})M{\o}ller, Safavimanesh \&
  Rasmussen]{moeller:safavimanesh:rasmusssen:15}
M{\o}ller J, Safavimanesh F, Rasmussen J. 2015{\natexlab{b}}.
The cylindrical {$K$}-function and {P}oisson line cluster point processes.
CSGB {R}esearch {R}eports~3, {Centre for Stochastic Geometry and Advanced
  Bioimaging}.
Preprint on arXiv: 1503.07423. To appear in Biometrika

\bibitem[M{\o}ller et~al.(1998)M{\o}ller, Syversveen \&
  Waagepetersen]{moeller:syversveen:waagepetersen:98}
M{\o}ller J, Syversveen AR, Waagepetersen RP. 1998.
Log {G}aussian {C}ox processes.
\textit{Scandinavian Journal of Statistics} 25:451--482

\bibitem[M{\o}ller \& Toftager(2014)]{moeller:toftager:14}
M{\o}ller J, Toftager H. 2014.
Geometric anisotropic spatial point pattern analysis and {C}ox processes.
\textit{Scandinavian Journal of Statistics} 41:414--435

\bibitem[M{\o}ller \& Torrisi(2005)]{moeller:torrisi:05}
M{\o}ller J, Torrisi G. 2005.
Generalised shot noise {C}ox processes.
\textit{Advances in Applied Probability} 37:48--74

\bibitem[M{\o}ller \& Waagepetersen(2004)]{moeller:waagepetersen:03}
M{\o}ller J, Waagepetersen RP. 2004.
Statistical inference and simulation for spatial point processes.
Chapman and Hall/CRC, Boca Raton

\bibitem[M{\o}ller \& Waagepetersen(2007)]{moeller:waagepetersen:07}
M{\o}ller J, Waagepetersen RP. 2007.
Modern statistics for spatial point processes.
\textit{Scandinavian Journal of Statistics} 34:643--684

\bibitem[Mountcastle(1957)]{mountcastle:57}
Mountcastle VB. 1957.
Modality and topographic properties of single neurons of cat's somatic sensory.
\textit{Journal of Neurophysiology} 20:408–434

\bibitem[Murray et~al.(2006)Murray, Ghahramani \& MacKay]{murray2006}
Murray I, Ghahramani Z, MacKay DJC. 2006.
{MCMC} for doubly-intractable distributions. In \textit{Proceedings of the 22nd
  Annual Conference on Uncertainty in Artificial Intelligence (UAI-06)}. AUAI
  Press,  359--366

\bibitem[{Myllym{\"a}ki} et~al.(2015){Myllym{\"a}ki}, {Mrkvicka}, {Grabarnik},
  {Seijo} \& {Hahn}]{myllymaki:etal:15}
{Myllym{\"a}ki} M, {Mrkvicka} T, {Grabarnik} P, {Seijo} H, {Hahn} U. 2015.
Global envelope tests for spatial processes.
Preprint on arxiv: 1307.0239v4

\bibitem[Nguyen \& Zessin(1979)]{nguyen:zessin:79b}
Nguyen X, Zessin H. 1979.
{Integral and differential characterizations of {G}ibbs processes}.
\textit{Mathematische Nachrichten} 88:105--115

\bibitem[Okabe \& Sugihara(2012)]{okabe:sugihara:12}
Okabe A, Sugihara K. 2012.
Spatial analysis along networks.
Wiley, Chichester

\bibitem[Proke{\v s}ov{\' a} et~al.(2014)Proke{\v s}ov{\' a}, Dvo{\v r}{\' a}k
  \& Jensen]{prokesova:dvorak:jensen:15}
Proke{\v s}ov{\' a} M, Dvo{\v r}{\' a}k J, Jensen E. 2014.
Two-step estimation procedures for inhomogeneous shot-noise {C}ox processes.
CSGB Research Reports~2, Centre for Stochastic Geometry and Advanced Bioimaging

\bibitem[Proke{\v s}ov{\' a} \& Jensen(2013)]{prokesova:jensen:13}
Proke{\v s}ov{\' a} M, Jensen EBV. 2013.
Asymptotic {P}alm likelihood theory for stationary point processes.
\textit{Annals of the Institute of Statistical Mathematics} 65:387--412

\bibitem[Rafati et~al.(2016)Rafati, Safavimanesh, Dorph-Petersen, Rasmussen,
  M{\o}ller \& Nyengaard]{AliEtAl:16}
Rafati AH, Safavimanesh F, Dorph-Petersen KA, Rasmussen JG, M{\o}ller J,
  Nyengaard J. 2016.
Detection and spatial characterization of minicolumnarity in the human cerebral
  cortex.
\textit{Journal of Microscopy} 261:115--126

\bibitem[Ripley(1977)]{ripley:77}
Ripley B. 1977.
Modelling spatial patterns (with discussion).
\textit{Journal of the Royal Statistical Society: Series B (Statistical
  Methodology)} 39:172--212

\bibitem[Robeson et~al.(2014)Robeson, Li \& Huang]{robeson:li:huang:14}
Robeson SM, Li A, Huang C. 2014.
Point-pattern analysis on the sphere.
\textit{Spatial Statistics} 10:76--86

\bibitem[Rue et~al.(2009)Rue, Martino \& Chopin]{rue:martino:chopin:09}
Rue H, Martino S, Chopin N. 2009.
Approximate {B}ayesian inference for latent {G}aussian models using integrated
  nested {L}aplace approximations (with discussion).
\textit{Journal of the Royal Statistical Society: Series B (Statistical
  Methodology)} 71:319--392

\bibitem[Ruelle(1969)]{ruelle:69}
Ruelle D. 1969.
Statistical mechanics: Rigorous results.
W.A. Benjamin, Reading, Massachusetts

\bibitem[Schoenberg(2005)]{schoenberg:05}
Schoenberg FP. 2005.
Consistent parametric estimation of the intensity of a spatial-temporal point
  process.
\textit{Journal of Statistical Planning and Inference} 128:79--93

\bibitem[Skare et~al.(2007)Skare, M{\o}ller \& Jensen]{skare:moeller:jensen:06}
Skare {\O}, M{\o}ller J, Jensen E. 2007.
Bayesian analysis of spatial point processes in the neighbourhood of {V}oronoi
  networks.
\textit{Statistics and Computing} 17:369--379

\bibitem[Stoyan(1979)]{stoyan:79}
Stoyan D. 1979.
Interrupted point processes.
\textit{Biometrical Journal} 21:607--610

\bibitem[Takacs(1986)]{takacs:86}
Takacs R. 1986.
Estimator for the pair-potential of a {G}ibbsian point process.
\textit{Statistics} 17:429--433

\bibitem[Tanaka et~al.(2008)Tanaka, Ogata \& Stoyan]{tanaka:ogata:stoyan:08}
Tanaka U, Ogata Y, Stoyan D. 2008.
Parameter estimation and model selection for {N}eyman-{S}cott point processes.
\textit{Biometrical Journal} 50:43--57

\bibitem[Taylor \& Diggle(2012)]{taylor:diggle:12}
Taylor BM, Diggle PJ. 2012.
{INLA} or {MCMC}? {A} tutorial and comparative evaluation for spatial
  prediction in log-{G}aussian {C}ox processes.
\textit{Journal of Statistical Computation and Simulation} 84:2266--2284

\bibitem[Valencia et~al.(2004)Valencia, Foster, Villa, Condit, Svenning
  et~al.]{valencia:etal:04}
Valencia R, Foster R, Villa G, Condit R, Svenning JC, et~al. 2004.
Tree species distributions and local habitat variation in the {A}mazon,
  {E}cuador: large forest plot in {E}astern {E}cuador.
\textit{Journal of Ecology} 92:214--229

\bibitem[{V}an~de Weygaert(1994)]{weygaert:94}
{V}an~de Weygaert R. 1994.
Fragmenting the {U}niverse {III}. {T}he construction and statistics of 3-{D}
  {V}oronoi tessellations.
\textit{Astronomy and Astrophysics} 283:361--406

\bibitem[{V}an~de Weygaert \& Icke(1989)]{weygaert:icke:89}
{V}an~de Weygaert R, Icke V. 1989.
Fragmenting the {U}niverse {II}. {V}oronoi vertices as {A}bell clusters.
\textit{Astronomy and Astrophysics} 213:1--9

\bibitem[{V}an Lieshout(2006{\natexlab{a}})]{lieshout:06a}
{V}an Lieshout MNM. 2006{\natexlab{a}}.
Campbell and moment measures for finite sequential spatial processes. In
  \textit{Proceedings Prague Stochastics 2006}, eds. M~Hu{\v s}kov{\' a},
  M~Jan{\v z}ura. Matfyzpress, Prague,  215--224

\bibitem[{V}an Lieshout(2006{\natexlab{b}})]{lieshout:06b}
{V}an Lieshout MNM. 2006{\natexlab{b}}.
Markovianity in space and time. In \textit{Dynamics \& Stochastics: Festschrift
  in Honour of M.S. Keane}, eds. D~Denteneer, F~{Den Hollander}, E~Verbitskiy.
  Institute for Mathematical Statistics, Beachwood,  154--168

\bibitem[{V}an Lieshout(2011)]{lieshout:11}
{V}an Lieshout MNM. 2011.
A {$J$}-function for inhomogeneous point processes.
\textit{Statistica Neerlandica} 65:183--201

\bibitem[{V}an Lieshout \& Baddeley(2002)]{lieshout:baddeley:02}
{V}an Lieshout MNM, Baddeley AJ. 2002.
Extrapolating and interpolating spatial patterns. In \textit{Spatial Cluster
  Modelling}, eds. AB~Lawson, D~Denison. Boca Raton, Florida: Chapman \&
  Hall/CRC Press,  61--86

\bibitem[Waagepetersen(2005)]{waagepetersen:05}
Waagepetersen R. 2005.
Discussion of `{R}esidual analysis for spatial point processes'.
\textit{Journal of the Royal Statistical Society,Series B} 67:662

\bibitem[Waagepetersen(2007)]{waagepetersen:07}
Waagepetersen R. 2007.
An estimating function approach to inference for inhomogeneous {N}eyman-{S}cott
  processes.
\textit{Biometrics} 63:252--258

\bibitem[Waagepetersen(2008)]{waagepetersen:08}
Waagepetersen R. 2008.
Estimating functions for inhomogeneous spatial point processes with incomplete
  covariate data.
\textit{Biometrika} 95:351--363

\bibitem[Waagepetersen \& Guan(2009)]{waagepetersen:guan:09}
Waagepetersen R, Guan Y. 2009.
Two-step estimation for inhomogeneous spatial point processes.
\textit{Journal of the Royal Statistical Society, Series B} 71:685--702

\bibitem[Waagepetersen et~al.(2016)Waagepetersen, Guan, Jalilian \&
  Mateu]{waagepetersen:guan:jalilian:mateu:16}
Waagepetersen R, Guan Y, Jalilian A, Mateu J. 2016.
Analysis of multispecies point patterns by using multivariate log-{G}aussian
  {C}ox processes.
\textit{Journal of the Royal Statistical Society: Series C (Applied
  Statistics)} 65:77--96

\bibitem[Walsch \& Raftery(2002)]{walsh:raftery:02}
Walsch DCI, Raftery AE. 2002.
Detecting mines in minefields with linear characteristics.
\textit{Technometrics} 44:34--44

\bibitem[Zhuang(2006)]{zhuang:06}
Zhuang J. 2006.
Second-order residual analysis of spatiotemporal point processes and
  applications in model evaluation.
\textit{Journal of the Royal Statistical Society, Series B} 68:635 --653

\bibitem[Zhuang(2015)]{zhuang:15}
Zhuang J. 2015.
Weighted likelihood estimators for point processes.
\textit{Spatial Statistics} 14:166--178

\end{thebibliography}

\end{document}